\def\galex{{\sl GALEX}}

\def\hst{{\sl HST}}
\def\acs{ACS}
\def\wfc3{WFC3}

\def\spitzer{{\sl Spitzer}}
\def\irac{IRAC}

\documentclass[12pt,preprint]{aastex}
\usepackage{epsf}
\usepackage{color}
\usepackage{graphicx}
\usepackage{graphics}
\usepackage{epsfig}
\usepackage{longtable}
\usepackage{multirow}
\begin{document}
\title{Photometric Evidence of an 
Intermediate-age Stellar Population in the Inner Bulge of M31}
\author{Hui Dong$^{1,2}$, Zhiyuan Li$^{3,4}$, Q. Daniel Wang$^{5,3}$, Tod R. 
  Lauer$^2$, Knut A. G. Olsen$^2$, Abhijit Saha$^2$, Julianne J. Dalcanton$^6$, Benjamin F. Williams$^6$}

\affil{$^1$ Instituto de Astrof\'{i}sica de Andaluc\'{i}a (CSIC), Glorieta de la Astronom\'{a} S/N, E-18008 Granada, Spain}\affil{$^2$ National Optical Astronomy Observatory,
Tucson, AZ, 85719, USA}\affil{$^3$ School of Astronomy and Space Science, Nanjing University, Nanjing, 210093, China}\affil{$^4$ Key Laboratory of Modern Astronomy and Astrophysics at Nanjing University, Ministry of Education, Nanjing 210093, China}\affil{$^5$ Department of Astronomy, University of Massachusetts,
Amherst, MA, 01003, USA}\affil{$^6$ Astronomy Department, University of Washington,
Seattle, WA, 98195, USA}\affil{E-mail: hdong@iaa.es}

\begin{abstract}
We explore the assembly history of the M31 bulge within a projected
major-axis radius of 180\arcsec\ ($\sim$680 pc) by studying its stellar populations in  
Hubble Space Telescope \wfc3\ and
\acs\ observations. Colors formed by 
comparing near-ultraviolet vs.\ optical bands are found to 
become bluer with increasing major-axis radius, which is 
opposite to that predicted if the sole sources of
near-ultraviolet light were old extreme horizontal branch 
stars with a negative radial gradient in metallicity. Spectral energy distribution fits require a metal-rich intermediate-age
stellar population 
(300 Myr to 1 Gyr old, $\sim$$Z_{\odot}$) in addition to the dominant
old population. The radial gradients in age and metallicity 
of the old stellar population are consistent with those in  
 previous works. For the intermediate-age population, we 
find an increase in age with radius and a mass fraction that increases 
up to 2\% at  680 pc away from the center. We
exclude contamination from the M31 disk and/or halo as the main
origin for this population. Our results thus suggest that intermediate-age 
stars exist beyond the central 
5\arcsec\ (19 pc) of M31 and contribute $\sim$1\% of the total 
stellar mass in the bulge. These stars could be related to the secular growth 
of the M31 bulge. 
\end{abstract}

\section{Introduction}



As the nearest massive spiral galaxy~\citep[$\sim$780 kpc,][]{mcc05}, 
the Andromeda galaxy (M31) provides an excellent laboratory with which 
 to study 
 the construction of galactic bulges. 
Compared to the bulge of the Milky Way (MW), 
the advantages of the M31 bulge lie in 
that i) stars are at a single, well-determined distance and 
ii) it is free of the strong, spatially non-uniform line-of-sight 
extinction against the Galactic bulge. Indeed, the M31 bulge is 
transparent across almost the entire electromagnetic spectrum, allowing 
for a panchromatic view of its stellar (and interstellar) 
components~\citep[][for a recent review]{li09}. 
These advantages largely reduce the uncertainty in deriving  
the intrinsic spectral energy distribution (SED) of the bulge, 
which is critical to determining the ages and metallicities  
of the underlying stellar populations. 

The stellar populations in the inner bulge of M31 are quite different  
from those of our Galaxy. Around the Galactic center, there are a 
number of giant molecular clouds ($\sim$4$\times 10^7~M_{\odot}$) and 
young massive stars~\citep{mor96}. In contrast, there are few signs of massive stars in 
the inner bulge of M31. 
\citet{ols06} investigate the near-infrared (NIR) color-magnitude diagrams (CMDs) of several bulge 
regions, $>$2\arcmin\ ($\sim$450 pc) away from M31*, the central super-massive 
black hole~\citep{dre88,kor88,cra92,gar10,li11}, and find that stars there are 
old ($>$6 Gyr) and metal-rich ($>$$Z_{\odot}$) \textbf{(see also~\citealt{ste03,sar05})}. 
~\citet{sag10} obtain 
long-slit optical spectra (4787 \AA$-$5857 \AA ) in the central 5\arcmin\ of the M31 bulge,
 and in most of these spectra find that multiple spectroscopic line 
indexes can be fitted with a single metal-rich 
stellar population of age around 12 Gyr old. Based on the ultraviolet (UV) CMD 
from {\sl Hubble Space Telescope} (\hst ) Faint Object Camera (FOC)
 observations of the central
14\arcsec$\times$14\arcsec\ (53$\times$53 pc) field of M31 in two bands (F175W and F275W),~\citet{bro98} suggest that 
detected UV 
point sources are not young massive stars, but mainly old 
evolved low-mass stars, a result confirmed over a larger area by~\citet{ros12}. 

 
These evolved low-mass stars, {\it i.e.,} core helium burning extreme 
horizontal branch (EHB) stars and their descendants, have high effective surface 
temperatures $T_{eff}>10^4$ K and can contribute substantially to the UV emission of the M31 
bulge. When evolved low-mass 
stars leave the 
EHB phase, they do not have significant stellar envelopes and thus do not 
evolve into the asymptotic giant branch (AGB) phase, and instead either leave the 
AGB phase early (PE-AGB) or fail to enter the phase entirely (AGB-manqu\'e). 
These hot-post HB stars (HP-HB stars) are relatively long-lived ($\sim$10 Myr,~\citealt{oco99}). 
In contrast, the stars with 
enough envelopes will complete the AGB phase and then the post-AGB (PAGB) phase. 
PAGB stars are more luminous than the PE-AGB and
AGB-manqu\'e stars, but because of their short 
lifetimes ($\sim$$10^4$ yr, see review by~\citealt{oco99}),  PAGB stars contribute little to the integrated emission 
 in the UV bands. 
Indeed, based on the latest \hst\ Wide Field Camera 3 (\wfc3) UVIS 
observations,~\citet{ros12} support the argument of~\citet{bro98} that most of the 
detected UV sources in the F275W and F336W bands 
should be HP-HB stars, which can 
account for only 2\% of 
the integrated bulge light in the F275W band.~\citet{bro98} and~\citet{ros12} suggest
that the remaining extended UV emission arises from the combination of 
numerous unresolved EHB stars, which must be present to explain the HP-HB populations  that 
is observed.~\citet{ros12} also find that the fraction of stars passing through this channel increases
 towards the center of the galaxy, though it is never more than 3\%. 
 They attribute this trend, as well as the positive gradient 
in the \galex\ FUV$-$NUV color distribution~\citep{thi05}, which are seen over 3.5 kpc, 
to the negative 
radial gradient in the stellar metallicity found by~\citet{sag10}, on the theory that 
 metallicity affects the strength of stellar winds and hence 
the mass of stellar envelopes and consequently the 
EHB and HP-HB populations. The FUV$-$NUV color 
measures the slope of the so-called UV upturn~\citep[][and references therein]{thi05,oco99}. 

The source of UV emission becomes more complicated at the very center of M31, 
where~\citet{lau98} suggest that the FUV peak at 
the dynamical center of M31 found by~\citet{kin95} is slightly 
extended and is not due to an AGN.~\citet{ben05} 
further find that the optical spectrum extracted from the central
0.3\arcsec\ field (1.1 pc) is consistent with a population of 
A-type stars or a ~200-Myr-old starburst. With observations from the \hst\
Advanced Camera for Surveys (\acs ) High-Resolution Channel 
(HRC),~\citet{lau12} resolve this central star cluster into twenty 
near-UV (NUV) bright stars and argue that they cannot 
be old stars in the PAGB phase, whose evolutionary duration
 is short.~\citet{sag10} also show that an old stellar population 
is not consistent with their derived  
spectroscopic line indexes within the central 5\arcsec\
(19 pc) region and that an intermediate-age stellar population is needed. 
The origin of these intermediate-age stars (several hundreds of 
Myr old) remains an enigma.~\citet{sag10} suggest that a gas-rich minor merger 
may have triggered a nuclear starburst. If this scenario is valid,
it is natural to ask whether or not the same event could 
have triggered star formation in a larger portion of the M31 bulge; how much 
stellar mass this event could have contributed to the bulge and  
what is the event's role on the evolution of the whole bulge. 


In this paper, we concentrate on studying the stellar 
populations in the inner bulge of M31 with the latest multi-band \hst\ dataset 
and using our results to constrain the assembly history of the M31 bulge. 
We utilize \hst\ \wfc3\ and
\acs\ observations in ten filters from
the near-UV to near-IR (listed in Table~\ref{t:obs}), which cover the central
250\arcsec$\times$250\arcsec\ ($\sim$950$\times$950 pc) region of M31. Previous
photometric observations obtained by ground-based telescopes in the 
optical/near-infrared bands 
suffer from confusion and differential extinction due to limited and varied resolutions and
sensitivities, which hamper the study 
of the spatial variation of underlying 
stellar populations. Images in our \hst\
dataset have superb resolution and sensitivity and cover a broad
 wavelength range. The inclusion of the 
 near-UV filters in particular is crucial to break
the age-metallicity degeneracy in the 
SED fitting. Uncertainties in derived ages and metallicities are smaller by more than 60\% , 
when the UV bands are included, than when only optical bands are used, based on studies  
of 42 globular clusters from~\citet{kav07}. These authors find that SED fitting 
using UV photometry also outperforms the spectroscopic index method. Further 
the intermediate-age stellar population are expected to have SEDs that peak at 4000 \AA , 
making the UV as important constraint. Using fits of this full spectrum 
coverage, we  find evidence for an intermediate-age stellar population. 



The reminder of the paper is organized as follows. We describe the observations and
data reduction in \S\ref{s:observation} and the color gradients 
along the minor axis of the M31 bulge in \S\ref{s:color}. 
Our method and analysis
procedure are presented in
\S\ref{s:method} and in \S\ref{s:analysis}, respectively. 
We discuss our results in \S\ref{s:discussion} and
summarize our conclusions in \S\ref{s:summarize}. 

\section{Observation and Data Reduction}\label{s:observation}
\subsection{\hst\ Multi-Wavelength Dataset}\label{ss:dataset}
The majority of our dataset comes from the
Panchromatic Hubble Andromeda Treasury~\citep[PHAT;][]{dal12,wil14}. 
As an \hst\ legacy program, this survey mapped roughly one third of the M31 disk
with the \hst\ \wfc3\ and \acs\ in six bands, F275W, F336W, F475W, F814W,
F110W and F160W. Except for the F110W
filter, various dither patterns (see Table~\ref{t:obs}) were
used to remove bad pixels and cosmic-rays, as well as to alleviate  
undersampling.
The PHAT data include the observations in both the shortest and
longest wavelength bands (F275W: 2700 \AA\ and F160W: 1.5 $\mu$m) used in this
work, giving important leverage needed to constrain extinction
and age. F336W and F475W on the blue and red 
sides of the 4000 \AA\ break provide a sensitive age 
measure of the underlying stellar populations. The F814W, F110W and F160W 
bands, being at longer wavelengths, 
 are insensitive to extinction and  
age and provide a useful color indicator of metallicity. 

We include observations taken in four additional UV/optical bands 
to increase our wavelength coverage. 
Three of them come
from Program GO-12174 (Li et al, in preparation): 
two medium bands (F390M,
 F547M) and one narrow band (F665N). Their exposure times
  are longer than those of the PHAT program. As a result, 
 signal-to-noise ratios (S/Ns) of these images are comparable to those of the 
wide filters used for PHAT. These additional single-pointing observations 
covered a limited 
area of sky. We use the F547M band as an analog of the
traditional Johnson $V$ band, because of their similar central
wavelengths. The fourth 
 filter, F435W, comes from observations taken as part of three 
different programs: Program 10006, 10760 and 11833~\citep{wil05}. 
These observations were 
originally taken to monitor the optical counterparts of
 X-ray novae in the M31 bulge. 
With various pointing directions, 
rotation angles and exposure times, the F435W 
observations covered roughly the field of interest, except for its southwest corner. 
The observations were all taken with dither patterns. Fig.~\ref{f:ehb} gives the low-resolution UV 
and optical spectrum of the central 10\arcsec$\times$20\arcsec\ 
region of M31 obtained from~\citet{oco99}, with overlaid the transmission curves of  
six UV and optical bands in our dataset, as well as those of two \galex\ bands; FUV and NUV, to 
demonstrate different spectral features covered by these filters. 

\subsection{Data Reduction}\label{ss:reduction}
The raw data and calibration files were downloaded from the Multimission Archive at STScI
(MAST). The basic calibration of each dithered exposure was
made with the latest \hst\ pipeline, {\tt OPUS}
 version 2010\_4, {\tt CALACS} version 5.1.1 and {\tt CALWFC3} version 2.1. The 
 steps included identifying bad pixels, bias correction, dark subtraction and flat
 fielding. 

The `{\tt Astrodrizzle}' task, based on {\tt PyRAF}\footnote{`{\tt PyRAF}' and `{\tt Astrodrizzle}' are the product of 
the Space Telescope Science Institute, which is operated by AURA for
NASA.}, was used to register individual dithered exposures,
correct for distortion, mask out defects (including cosmic rays), and
combine the dithered exposures into pointing images. 
While `{\tt Astrodrizzle}' only corrected for the distortion of the F110W
images, cosmic-rays in individual F110W images were  
identified in the {\tt CALWFC3} step, through the use of the 
sequential readouts
`{\tt MULTIACCUM}' mode. For the F275W and F336W observations,
in which the \wfc3/UVIS chip gap was covered by only one dithered
exposure, cosmic rays could not be readily identified. 
For simplicity, we manually removed the regions covered by the gap before 
mosaicking the images. 

The construction of the mosaic 
for each filter follows a few steps. First, for the F475W band, we used the technique presented
in~\citet{don11} to correct for the astrometry and bias offset among
different pointings. 
We tied our images to the absolute astrometric frame defined by 
the 2MASS catalog~\citep{skr06}. Second, we corrected for 
the astrometry of any of the near-UV or optical image relative to 
the F475W band, using commonly detected stars. 
We aligned the near-IR F110W or F160W images to the 
astrometry-corrected F814W images (which were themselves 
aligned to the F475W images and have more stars 
in common with the near-IR bands). 
The bias offsets among different pointings in these nine filters were
also removed. Finally, we merged the images to form the mosaics in the ten bands. 
These mosaics cover a 
central 250\arcsec$\times250$\arcsec\ ($\sim$950$\times$950 pc) field of M31. 
The `PHOTFLAM' values listed in
Table~\ref{t:obs} are used to convert the image units from counts
(electron/s) to flux density (erg cm$^{-2}$ s$^{-1}$ \AA $^{-1}$). 


Well-characterized uncertainties are crucial for 
the SED fitting. The photometric uncertainty 
at each pixel in a mosaiced image ($\sigma_n$; the subscript stands for the n$^{th}$ filter) 
consists of three main factors: 1) the
Poisson statistical noise of the electronic signal; 
2) the error in the bias and flat-field calibration; and
3) the systematic
uncertainty in the photometry, as described by the fits header 
`PHOTFLAM' keyword. 
We empirically characterize the factors  
1) and 2), using the method described in~\citet{don14}. Briefly, 
for each pixel, we calculated the median and 68\% percentile 
of fluxes in its adjacent $0.65\arcsec\times0.65\arcsec$ field 
as the estimates of the local background and its uncertainty. 
The median ratio of the statistical uncertainties to the fluxes of
individual pixels in each of the
ten-band mosaics is listed in Table~\ref{t:obs}.
The WFC3 and ACS systematic calibration errors\footnote{WFC3:
  http://www.stsci.edu/hst/wfc3/phot\_zp\_lbn and ACS: Bohlin, R.,
  2012, acs, rept, 1B. Private communication with STScI helpdesk} are
$\sim$2\%. We found that with the same
incoming spectrum, the fluxes derived from the {\tt SYNPHOT} 
in Chips 1 and 2 of \wfc3/UVIS in the F275W band could differ by
up to 3\%\footnote{The difference could be due to `{\tt SYNPHOT}',
  \hst/\wfc3\ STAN Issue 11}. Therefore, in \S\ref{s:analysis}, we 
adopt the mean value of the F275W fluxes
from the {\tt SYNPHOT} for Chips 1 and 2 of \wfc3/UVIS, including
 an extra 1.5\% systematic error to the F275W band during 
the SED fitting. The systematic errors due to the factor 3) are also listed in
Table~\ref{t:obs}. In \S\ref{s:analysis}, when we present the SED fitting 
for a selected region, where the integrated flux uncertainty in each band is 
the square root of the quadratic sum of the empirical statistic errors in 
all the included 
pixels in this region, together with the systematic errors.

We corrected for foreground Galactic extinction~\citep[E(B$-$V)=0.062, 
{\it i.e.,} $A_V$=0.17,][]{sch11} for all bands. The
relative extinctions ($A_{n}/A_{F547M}$, where 
$A_{n}$ or $A_{F547M}$ is the absolute
extinction in the $n^{th}$ or F547M band) derived from the MW-type
extinction law in~\citet{fit99} for an old metal-rich stellar
population (12 Gyr and solar metallicity) are listed in 
Table~\ref{t:obs}. As discussed in the Appendix A
of~\citet{don14}, the relative extinctions
($A_{n}/A_{F547M}$) in the ten \hst\ bands are insensitive to 
the adopted background spectrum. 

\subsection{Region Selection}\label{ss:region}
To break the well-known degeneracy between extinction, age 
and metallicity, we selected only low extinction regions 
for the present study. Fig.~\ref{f:ratio} presents a map of NIR-to-NUV 
(F160W/F336W) intensity ratio for the 
inner bulge.  This ratio increases with the foreground 
extinction, as well as with the age and metallicity of unresolved stars. 
Fig.~\ref{f:ratio} shows that the ratio is  
generally high in regions with hot dust, as traced by the \spitzer/IRAC 
`dust-only' 8 $\mu$m intensity map of~\citet{li09}. This conclusion breaks down in the
central 10\arcsec\ ($\sim$38 pc) region, where there is a high 
ratio of F160W to F336W, but no corresponding 
excess in 8 $\mu$m emission. We discuss this region in \S\ref{ss:under}.
 A previously known dusty clump located at 30\arcsec\ ($\sim$120 pc) 
southeast of M31* (as outlined by the blue ellipse in Fig.~\ref{f:ratio}) has 
significant 8 $\mu$m emission and is also detected in
CO observations~\citep{mel13}, but is not prominent in our
flux ratio map;~\citet{mel13} suggest that this dusty clump
is located at the far side of the bulge.

The mean ratio in the South East quadrant of Fig.~\ref{f:ratio} is smaller than those 
in the other three quadrants by at least 10\%, indicating low 
foreground extinction. We isolate two sectors in this region
 to avoid the dusty 
clumps apparent in the flux ratio map and in the 
\spitzer/IRAC 8 $\mu$m observations. We use the region between 
PA = 110$^o$ to 180$^o$ (green sector) for the central 
100\arcsec , while beyond this radius, we use 
 the region between PA=150$^o$ to 180$^o$ (blue sector), to
avoid a dusty clump at PA$\sim$140$^o$ and 160\arcsec\ in projected
radius away from M31*.  
According to Li et al. (in preparation), the
two-dimensional light distribution in the M31
bulge in the ten \hst\ bands can be characterized by elliptical isophotes with a 
PA=50.9$^o$ (East of North) and an axis ratio of 0.8. We divide 
these two chosen sectors into elliptical annuli steps of 5\arcsec\ 
along the major-axis for the SED fitting in \S\ref{ss:under}. We will refer to ``radius'' in 
this paper as being major axis length of each ellipse. 

\section{Color Gradients}\label{s:color}
We first analyze the radial gradients in UV-optical colors, which 
give us hints about the spatial variation of the 
properties of underlying stellar populations.   
Fig.~\ref{f:surface_dis} presents the \hst\ \wfc3/\acs\ colors obtained 
from the light intensities in four 
bands and along the minor-axis of the M31 bulge out to 144\arcsec , 
corresponding to a major-axis 
distance of 180\arcsec\ (680 pc) and restricted to the two sectors we are using for analysis.  
In the same plot, we overlay the radial surface brightness distribution 
of the \spitzer/\irac\ `dust-only' 8 $\mu$m emission. The colors tend to become 
bluer when moving away
from M31* and are not correlated with the dust distribution. Therefore, the
color gradients cannot be solely caused by
the decrease of extinction 
and most likely represent a change in the intrinsic properties of the stellar 
populations. 

Our \textbf{\hst} UV-optical color gradients differ from those of \galex\ FUV$-$NUV in~\citet{thi05}. 
Fig.~\ref{f:galex_hst} shows the radial color distributions 
of: F275W$-$F475W and \galex/FUV$-$F475W, as well as  \galex\ 
FUV$-$NUV. If the integrated 
flux in the F275W band was dominated by 
the EHB stars, as suggested by~\citet{ros12}, the F275W$-$F475W color, just like the FUV$-$F475W color, 
can then be used to constrain the ratio of EHB to 
main-sequence turnoff stars. We would expect that the 
FUV$-$F475W and F275W$-$F475W colors follow the FUV$-$NUV 
color, becoming increasingly red with the galactocentric distance, 
due to the decreasing contribution from EHB stars. But in the fact, the 
F275W$-$F475W color turns bluer when moving away from M31*; 
although the FUV$-$F475W color shows a positive radial gradient within 
the central $\sim$50\arcsec\ (190 pc), it turns negative beyond that. These different 
color gradients suggest that the same population of stars that dominates the FUV flux 
(presumed by EHB stars in the old stellar population) probably dominate the FUV channel, but 
do not dominate the intensity in F275W 
and that instead the variation of the F275W$-$F475W color along the radius may be due 
to another stellar population (see further 
discussion in \S\ref{sss:evolved}). 

\section{The SED-fitting Method}\label{s:method}
We compare the observed and theoretical SEDs to constrain the properties
of underlying stellar populations. To do this, we minimize the deviation between the observed
flux in the n$^{th}$ filter, $I_n$, and the $S_n$ predicted 
by a stellar synthesis model of given stellar age and metallicity, 
\begin{equation}\label{e:chi}
\chi^2=\sum^{N}_{n=1}\frac{\{
  I_n-[(1-f)+f\times10^{-0.4\times\frac{A_{n}}{A_{F547M}}A_{F547M}}]S_n\aleph\}^2}{\sigma_n^2}
\end{equation}
where $f$, $A_{n}$/$A_{F547M}$, $A_{F547M}$ and $\aleph$ are the fraction of obscured starlight, 
the relative extinction, the absolute extinction and the normalization, respectively.  $\sigma_n$ is the
uncertainty of $I_n$, as described in \S\ref{ss:reduction}. The sum in this expression is over all 
available filters (N$\leq$10).

We utilize `Starburst99' ~\citep{vaz05} to
calculate the theoretical SEDs. Padova stellar synthesis models with full
AGB treatments (but not the EHB stars and their descendants) are adopted. We use  the
default~\citet{kro02} Initial Mass Function (IMF) in Starburst99 
($d\Phi\propto M^{-\Gamma}dM$, $\Gamma$=1.3
for 0.1 M$_{\odot}$$<$$M$$<$0.5 M$_{\odot}$, 2.3 for
0.5 M$_{\odot}$$<$$M$$<$100 M$_{\odot}$).~\citet{con12} 
suggest that the IMF of the nuclear region of M31 is normal, rather than  
bottom heavy. 
While `Starburst99' does not
include the recipe for EHB stars, the presence of such stars 
should not affect our results, because their contribution in our 
multi-bands is small (see \S\ref{s:color} and \S\ref{sss:evolved}). We choose 
instantaneous burst populations of ages from 100 Myr
to 15 Gyr with a step size of 0.01 dex (Myr) and five default
metallicities (Z=0.02 $Z_{\odot}$, 0.2 $Z_{\odot}$, 0.4 $Z_{\odot}$,
1 $Z_{\odot}$ and 2.5 $Z_{\odot}$). We generate spectra for these ages
and metallicities, which are then convolved with the filter transmission 
curves using the `{\tt SYNPHOT}' to obtain the corresponding fluxes in  
the ten \hst\ bands. 
We then linearly interpolate the fluxes from these grids to obtain the
fluxes of stellar populations with various ages (from 100 Myr to
15 Gyr) and metallicities
(from 0.02 $Z_{\odot}$ to 2.5 $Z_{\odot}$). 

Before we perform the $\chi^2$ minimization,
 we choose to fix two parameters. The first parameter is the fraction of the 
 starlight that is obscured by dust in M31, 0$<$$f$$<$1, which is 
 degenerate with $A_n$~\citep[see Fig. 1 of][]{don14}. 
However, $A_n$ is  
insensitive to $f$ in regions of low attenuation. This applies to 
the region that we select for the SED-fitting 
in \S\ref{ss:region}. Therefore, we
simply fix $f$=1 ({\it i.e.,} assuming that extinction exclusively arises from the
 foreground of the bulge). 
Setting $f$ free in the SED-fitting would not
change the best-fit parameters, but would increase the uncertainties of the output ages 
and metallicities. The second parameter is the
relative extinction, $A_n$/$A_{F547M}$. We adopt the average
extinction curve ($A_{n}$/$A_{F547M}$) derived from five dusty clumps within the central
1\arcmin\ ($\sim$230 pc) radius of M31 (\citealt{don14}; Table~\ref{t:obs}), which is characterized 
by $R_V$=2.4. We also test MW-type extinction curve in \S\ref{s:analysis} and our results 
remain unchanged, again because of low foreground extinction to the regions of interest.   


\section{Results}\label{s:analysis}

\subsection{The Presence of Intermediate-age Stars}\label{ss:two}

As an illustration of our method, we first select a region in the inner bulge of M31 (blue box in Fig.~\ref{f:ratio}) to 
examine whether or not an intermediate-age stellar population may 
exist beyond the central 5\arcsec\ (19 pc) of the M31 bulge. 
This region (5\arcsec$\times$5\arcsec , 19 pc$\times$19 pc) is
$\sim$11\arcsec\ (42 pc in projection) southeast of M31* and appears dim in the
\spitzer/\irac\ `dust-only' 8 $\mu$m image, indicating little local foreground extinction. The size of the 
region is large enough for us to obtain the photometry with statistical  
uncertainties $<$0.3\% in all the ten bands and is sufficiently 
small to neglect any potential variation of the SED along the minor-axis of
the bulge (\S\ref{ss:under}). 

We first fit the observed SED with one
instantaneous stellar population as suggested by~\citet{sag10}. This fit has four free 
parameters: $A_{F547M}$, age,
metallicity and 
normalization ($\aleph$). We search for the best fit with 
{\tt MPFIT}~\citep{mar09}. The fit is rather poor, especially for the UV 
bands (see Fig.~\ref{f:fit_com}), 
with $\chi^2$/d.o.f.=24.6/6=4.1. The age and metallicity of the
best-fit single stellar
population are $\sim$4.4 Gyr and 2.2 Z$_{\odot}$, respectively, with 
$A_{F547M}$$\sim$0.08. 
The age is much smaller than 12 Gyr as derived by~\citet{sag10}. 
If we fix the metallicity to the value of
[Z/H]=0.16 (1.4 $Z_{\odot}$) derived
from Fig.12 of~\citet{sag10} at the corresponding radius, the reduced $\chi^2$ becomes even larger and the
fitting is still poor for the UV bands. 

We thus add another instantaneous stellar population to try to improve the fit. 
The model now includes three more parameters: the age,
metallicity and mass fraction of the second stellar population. The resultant 
$\chi^2$/d.o.f. of the model fit is $\sim$ 4.8/3=1.6, which is a 
significant improvement over the fit with the single population. According to 
the $F$-test, the fit is improved
with the null-hypothesis probability $p$=0.13. The fitted extinction is 
formally zero, consistent 
with the ratio map in
Fig.~\ref{f:ratio}.  The new 
model fits the UV bands well (see Fig.~\ref{f:fit_com}). The age 
of the old stellar population becomes 14.8 Gyr\footnote{This value is older than 
the Universe's age, 13.8 Gyr. However, considering its uncertainty, 0.15 dex, it does 
not contradict the scenario that these stars formed in 
the early Universe.}, while the metallicity is
$\sim$1.5 $Z_{\odot}$, similar to the values obtained by ~\citet{sag10}. The second 
stellar population has an age of  $\sim$660 Myr and a metallicity of 2 $Z_{\odot}$. 
We estimate the uncertainties of the fitting parameters via a Monte Carlo 
simulation. We generate 100 fake SEDs based on the best-fit model  
and the flux measurement errors (assumed to be 
Gaussian; \S\ref{ss:reduction}). 
Each of the fake SED is refitted. The 68\% percentiles around  the 
best fit value of a parameter are used as its 1$\sigma$ uncertainty. 
\textbf{Because the SED predicted by the stellar synthesis model is a non-linear 
function of the input parameters, these errors are not symmetric to the best-fit 
value.} In Fig~\ref{f:ya_ym}, we show the correlation between
the age and metallicity of the second stellar 
population. 
The age is tightly constrained ($\sim$0.15 dex), whereas 
the metallicity has a large uncertainty
($\delta$[Z/H]$\sim$0.4). This 
intermediate-age stellar population, though accounting for only
$\sim$0.9\% of the total mass, contributes more than
50\%, 20\% and 25\% of the integrated intensity in the F275W, 
F336W and F390M bands, respectively. 

We also perform three tests to check the reliability of our result. First, we 
randomly select several regions 
of similar sizes and 
radius in the southeast of the M31 bulge and obtain
similar ages/metallicities for the two stellar
populations. Next, 
we check for EHB contamination in the F275W band. 
We repeat the two component fit excluding F275W band and find that 
the ages and the metallicities, as well as the light contribution by the
intermediate-age stellar population, vary by less than 20\%. Finally, we run the SED fitting without the 
UV bands (F275W, F336W, F390M). 
For a single instantaneous stellar population, the $\chi^2$/d.o.f. value 
reduces to 3.9/3=1.3. The resultant age and metallicity of 
9.8 Gyr and $\sim$Z$_{\odot}$ are similar to
those of~\citet{sag10}.  When we utilize two stellar populations, the
mass fraction of the intermediate-age stars decreases by a factor of
$\sim$10, which indicates that the optical and IR data alone give 
little indication for the presence of  the intermediate-age stellar population. Therefore, 
we conclude that our finding of the intermediate-age stellar population is firm.


\subsection{Spatial Variation of the Stellar
  Populations}\label{ss:under}
We now turn to investigate the radial variation of the stellar
populations in the inner bulge of M31. We apply the same SED fitting as in 
\S\ref{ss:two} for the annuli defined in \S\ref{ss:region} with two instantaneous starbursts: one traces the 
old stellar population and the other represents a younger population. Within the sector outlined in 
green (see Fig.~\ref{f:ratio}), data in all ten bands are available 
for the SED fit, for which the stellar ages and metallicities
of the two populations are left as free parameters. For the large off-center sector 
(100\arcsec to 180\arcsec , marked blue in Fig.~\ref{f:ratio}), which has data in 
only seven bands (six PHAT bands + F435W), we fix the age and
metallicity of the old stellar population to the best-fit values obtained for the annulus of 
95\arcsec $-$ 100\arcsec\ in the green sector, considering that from 100\arcsec\ to
180\arcsec , the properties of the old stellar population are likely
to be constant 
(see Fig.12 in~\citealt{sag10}). $A_{F547M}$ is also 
fixed to zero, which has little effect on the fit. 
The similar Monte Carlo simulations as used in \S\ref{ss:two} give
the uncertainties of these parameters, which are dominated 
by the systematic photometric errors in the ten bands (the statistical uncertainty in each annulus is small, $<$0.6\%). 
The fitting results are listed in
Table~\ref{t:annu_fit} and the radial profiles of the fitted 
parameters are plotted in Fig.~\ref{f:annu_chi}. If the two 
populations are distributed axisymmetrically in the bulge, the total mass of the old stellar population is 
1.4$\times$10$^{10}$ M$_{\odot}$ and the mass of the intermediated age population is  
1.5$\times$10$^8$ M$_{\odot}$ within 
the central 180\arcsec\ (680 pc).


\section{Discussion}\label{s:discussion}
Using a relatively extinction-free region in the inner bulge of M31, 
we have characterized the radial variations in age and metallicity of the stellar populations. Due to
the limitations of the available filters, we use only two
instantaneous starbursts to fit the observed SED. We have found 
strong evidence for the presence of an intermediate-age stellar population, 
in addition to a plethora of old stars in the inner bulge of the galaxy.

In this section, we first examine the possibility that the extra UV light, which 
has been attributed to the intermediate-age stellar population, may instead 
come from either old metal-rich, evolved low-mass stars in the  
bulge or main-sequence turnoff stars in the projected galactic disk and halo 
 (\S\ref{ss:f275w}). 
We then discuss why the intermediate-age stellar population has not been 
revealed in previous studies (\S\ref{ss:comp}). Lastly, we explore the potential origin of the  
population and the implications for 
the star formation history of the M31 bulge in \S\ref{ss:inter}. 

\subsection{Potential UV sources}\label{ss:f275w}
\subsubsection{Evolved Low-mass Stars}\label{sss:evolved}


We find that EHB stars cannot dominate the 
unresolved emission in the three UV bands: F275W, F336W and F390M. 
The emission peak of the coolest EHB stars with
temperature $\sim10^{4.15}$ K~\citep{oco99} is around 2100 \AA ,
near the short wavelength edge of the
transmission curve of the F275W filter. Furthermore, the intensity observed in
 the broad-band F275W image is contaminated by the
emission at optical wavelengths, where the M31 bulge is particularly bright, due 
to a well known red leak in the filter.  
\textbf{This contribution from optical light means that only a 
fraction of the observed F275W emission is available to potentially 
be explained by EHB stars.  The longer-wavelength F336W and F390M 
bands sit on the Rayleigh-Jeans tail of the EHB SED, 
making EHB stars very inefficient contributors to these bands.}
Our finding is 
supported by the data. In \S\ref{s:color}, we find that the gradients of 
F275W$-$F475W and FUV$-$NUV are totally different, suggesting that the 
FUV band is the only one where the EHB flux dominates. 
Further evidence is present in the UV-optical spectrum of the central 
10\arcsec$\times$20\arcsec\ region of 
M31 from~\citet{oco99} (Fig.~\ref{f:ehb}). The spectral intensity 
decreases from the optical toward the UV and then
 increases shortward of 2000 \AA , presumably 
due to the presence of EHB stars. We use a blackbody to fit the
spectrum below 2000 \AA\ minimizing the potential contamination from low-mass 
main-sequence turnoff stars and find that log (T$_{eff}$)$\sim$4.45, 
consistent with the value expected for EHB stars. From convolving the
UV-optical spectrum and the blackbody with the \hst\ \wfc3/UVIS
transmission curves, we find that the EHB contributes \textbf{at most}  
roughly 23\%, 2.9\% and 1.2\% of the total fluxes in the F275W, 
F336W and F390M bands, respectively, even in these central regions where the fraction 
of EHB stars is expected to be highest. These values 
decrease with increasing radius, because of 
a decreasing fraction of EHB stars due to the presumably decreasing metallicity. 

We suggest that main-sequence turnoff stars in the
intermediate-age stellar population found in
\S\ref{ss:two} explain the radial gradients of the near-UV and optical
colors. 
In \S\ref{ss:under}, we
find that this relatively young population contributes 
at least 40\% of the
emission in F275W (the rest is from numerous low-mass main-sequence 
turnoff stars in the old stellar population) in the central 5\arcsec . This contribution reaches $\sim$62\% at 180\arcsec\
away from M31*. The metallicity of this population seems to be constant beyond 30\arcsec .
Therefore, the increasing mass fraction of 
the population at larger radius is expected to cause the systematic blueing 
of the F275W$-$F475W color with increasing radius. In the central 50\arcsec , EHB stars probably 
still dominate the FUV and NUV emission, which explains the positive 
radial gradient in FUV$-$F475W. However, beyond the 
central 50\arcsec , the EHB contribution decreases while the 
FUV and NUV intensities are increasingly contributed by the 
intermediate-age stellar population. Fig.~\ref{f:fuv_int_galex} shows the radial distribution of the light 
contribution in the observed \galex\ FUV band for the intermediate-age stellar population. Although with 
large variation due to the limitation of our SED fitting, there 
is still a potential trend in this plot; the light contribution by this population increases from 
$\sim$20\% near M31* to $\sim$60\% at 180\arcsec . Therefore, the emission from 
main-sequence turnoff stars 
in that population offsets the metallicity effect of the old stellar population on 
FUV$-$NUV. As a result,  FUV$-$NUV increases much slowly with radius, compared to the central
50\arcsec . Thus, the intermediate-age stellar population detected here is in addition to 
the EHB stars suggested in previous works.

\subsubsection{Contamination by Disk and Halo Stars}
Considering that the mass fraction of the intermediate-age stellar population 
increases with radius, one might suspect 
a substantial contamination from stars in the disk 
and/or halo of M31. We quantify the contamination based on existing disk and halo models.~\citet{kor99} and~\citet{cou11} 
have decomposed the light contributions from the disk, bulge 
and halo of M31 through fitting the surface brightness distributions in the 
$V$ ad $I$ bands, 
respectively. Their decomposition gives the half-light radius of the
bulge and the scale length of the exponential disk as 0.8 kpc
(210\arcsec ) and 5 kpc (1320\arcsec ) (see also the 
recent characterization in~\citealt{dor13}). The scale 
length of the disk is substantially larger
than the size of our regions. With the small field-of-view 
considered here, their absolute contamination should be 
approximately uniform. Fig.~\ref{f:disk_con} 
compares the model-predicted contribution in the $V$ band from the bulge
and the disk and our measured contribution from intermediate-age stellar population. 
We find that within the central
$\sim$100\arcsec , the disk can account for only 40\% 
(16\%$-$49\% at the 68
percentile uncertainties) of the flux of 
the intermediate-age stellar population. Similarly, Figure 9 of~\citet{cou11} shows
that 
the halo contributes less
than 2\% in the $I$ band (the analog of the F814W band) in the inner bulge of M31. If halo stars are 13 Gyr old
and 0.02$Z_{\odot}$~\citep{kal06}, for example, we then expect that they contribute 
$\lesssim$ 3\% of the flux of the intermediate-age stellar population in the $V$ band. 
 Halo stars are also too
metal-poor to explain the intermediate-age stars. Most importantly, the observed 
mass surface density distribution of the intermediate-age stellar population (Fig.~\ref{f:disk_mass}) is very much different 
from a flat distribution as may be  
expected from the disk and halo contamination. Therefore, we conclude that the bulk of the intermediate-age stellar population   
indeed arises from the M31 bulge. 

\subsection{Comparison with Previous Work}\label{ss:comp}
Previous studies of the stellar populations in the inner bulge 
of M31 use optical and/or 
near-IR data which are not particularly sensitive to intermediate-age stars. 
\citet{ols06} analyze the images taken by Gemini/NIRI and 
\hst/NICMOS observations of the M31 bulge. Even with the
high angular resolution provided by the the Gemini/NIRI with Adaptive
Optics ($\sim$0.09\arcsec
) and \hst/NICMOS ($\sim$0.185\arcsec ), they detect only individual stars at 
the tip of the red giant branch (RGB) with ages $>$1 Gyr.~\citet{sag10} fit the 
ground-based long-slit optical spectra within the central 5\arcsec\ (19 pc) radius  
with two stellar populations. They find an intermediate-age stellar population with age $\sim$600 Myr, which contributes less than 10\% of the total
stellar mass, with the rest from the old stellar population ($\sim$8 Gyr), which is consistent with our result.  At 
larger radii, they only use one old stellar
population to infer the presence of a negative radial gradient in metallicity, as well as the stellar age of $\sim$12 Gyr, a trend we also 
see from the broad band fits.  They
then compare their Lick index analysis results with the u-g color of the SDSS data and conclude that they are consistent. 

 
The non-detection of the intermediate-age stellar population   
in bulge regions beyond the central 
5\arcsec\ in previous works is likely due to their detection limit. 
According to the Padova stellar evolutionary
tracks~\citep{bre12}, for a stellar population with Z$_{\odot}$ and
age=0.7$-$ 1 Gyr, stars with masses $>$3 M$_{\odot}$ have died, 
while stars
with 2.1$-$2.4 M$_{\odot}$ should have just left the main-sequence. 
The F275W and F336W
magnitudes of these latter stars ($>$25 mag) are still fainter than the 90\%
completeness limit of the PHAT survey in the UV band 
($\sim$24.5 at F275W and
F336W,~\citealt{ros12}). Their age can also explain why they are not detected in 
the study by~\citet{sag10}. In the Lick index method, the 
H$\beta$ 4861 \AA\ absorption line is a critical age discriminator, because it  
is strong in A-type stars of age around 
few hundred years. Under-subtracting the H$\beta$ emission line 
from the nuclear gas spiral surrounding M31*~\citep[]{li09} may have led to an 
overestimate of the stellar age based on the Lick index method, although~\citet{sag10} 
mention that they estimate the  properties of the underlying stellar population 
and emission lines, simultaneously. 
Because of the strong H$\beta$ absorption line, the existence of $\sim$300$-$600 
Myr old stars near M31* is well recognized. Away from M31*, the 
age of the intermediate-age stellar population detected here is large enough that the corresponding decrease in the
equivalent width of the H$\beta$ line makes it hard to detect the
 population in the integrated spectrum, 
which is dominated by the old stellar population. In \S\ref{ss:two}, 
we have shown that without the three UV bands, the
observed SED could indeed be fitted reasonably well with a single stellar population, reaching
 the same results as those of~\citet{sag10}. 
Therefore, 
including the UV bands into the SED fitting is critical to 
the identification of the intermediate-age stellar population. 
As a result, we are more sensitive to this population than any previous study.


\subsection{The Building History of the M31 Bulge}\label{ss:inter}
The existence of metal-rich intermediate-age stars in the M31 inner bulge is
not a unique phenomenon.~\citet{ben13} present a study of 
the high-resolution spectra of
58 dwarfs and supergiants in the Galactic bulge. Metal-poor stars are shown to have relatively old ages (10$-$12 Gyr), whereas metal-rich
ones ([Fe/H]$>$-0.1) have a wide range of ages from 2 to 12 Gyr and   
at least 5\% of them are younger than 5 Gyr. 
Based on a study of eight nearby spiral galaxies,
~\citet{mac09} find positive radial gradients of 
luminosity-weight age, which 
indicate the existence of young stellar populations in 
the galactic nuclei. 
They suggest that relatively young 
($<$ 1 Gyr) stellar
populations contribute as much as 70\% of the optical
emission from the galatic bulges, 
although their mass contribution are small, $<$20\%. 

The radial gradients in the age and metallicity of stellar populations provide 
insights into not only their origins, 
but the formation history of the M31 bulge as well. The stellar bulges of spiral galaxies 
may be classified into three types: 1) `classical', 2) `disk-like' and 3) boxy/peanut-shaped~\citep{kor04,ath05}. 
Classical bulges have S\`ersic index $>$2 and have properties similar to elliptical galaxies. 
They also fall into the fundamental plane of elliptical galaxies and are 
supported primarily by random motion~\citep{kor04}. 
Such spheroids are suggested to be constructed through monolithic 
collapses of primordial gas clouds and/or major mergers of galaxies~\citep[][and reference therein]{mac09}. In the monolithic 
collapse scenario, inner parts of spheroids experience 
an intense star formation in the early Universe over a short time interval. 
Heavy elements released by this activity are locked in the galactic nucleus, because 
of its deep potential well. As a result, stars formed in subsequent  
star formation tend to be  metal-rich. Meanwhile, metal-poor 
intergalactic gas could keep falling into the galaxy, which may trigger new star formation in outskirts 
and produce younger and metal-poorer stars. 
Therefore, this scenario predicts negative age and metallicity gradients. After a  
major merger, galactic disk could be disturbed, or even destroyed to form galactic bulge. 
The age and metallicity distributions in the new spheroid are smoothed and 
become flat. `Disk-like' bulges with S\`ersic index $<$2 are offen identified as `pseudobulges'. 
Boxy/peanut bulges are generally interpreted as edge-on barred 
pseudobulges~\citep{bur05}. A pseudobulge is likely built 
through `secular evolution'~\citep{kor04}. The disk instability introduces a 
rotating bar, which buckles and heats the disk vertically to increase its
scale height. The bar can also transport
the disk material into the inner region, leading to star formation and thus the growth of the bulge. 
This kind of bulge still keeps the memory 
of the galactic disk, so that they are primarily rotationally supported. Because it 
takes a long time for materials to spiral into the inner parts of such a bulge, a negative age 
gradient and a uniform metallicity are expected.

The M31 bulge is classified as a classical bulge, because of its S\`ersic index, 
2.2~\citep{cou11}, although this value is close to the boundary to a pseudobulge~\citep{fis10} and~\citet{bea07} 
finds boxy structures. The 
large velocity dispersion, $\sim$160 km/s~\citep{sag10}, also supports that M31 has 
a classical bulge. While the monolithic collapse scenario  
could explain the age and metallicity gradients of the old stellar population 
found in \S\ref{ss:under}~\citep[see also][]{sag10}, the supersolar metallicity 
indicates subsequent major merger(s). But the 
presence of the dynamic fragile galactic disk indicates that no recent major merger has happened in M31. 
Therefore, the high stellar metallicity and its negative gradient  
in the M31 bulge 
suggest that the early M31 experienced several major mergers, which 
triggered the formation of stars with enhanced metallicity. After that, accreted 
intergalactic materials have  
reduced the metallicity of stars in the outer parts of the bulge. 

On the other hand, the intermediate-age stars may be due to secular 
evolution, instead of mergers. M31 has likely experienced minor 
mergers recently, with its satellite galaxies, as indicated in stellar number density maps
produced by wide-field imaging surveys~\citep[][and references
therein]{iba14}. A potential head-on collision between M31 and 
M32 about 200 Myr ago has also been proposed to explain the young stars in the 
central 5\arcsec\ \citep[e.g.][]{sag10,lau12}, as well as the 
10 kpc star forming ring~\citep{blo06}; although recently
\citet{die14} suggest that head-on collision is not required 
to produce the ring and \citet{lew15} find that the 10 kpc ring is 
long-lived and stationary, which cannot be due to a purely 
collisional origin.~\citet{blo06} suggest that the M32 just 
passed the inner bulge of M31 recently, which triggered the 
star formation in its nucleus. This scenario predicts  
that the age of the intermediate-age stellar population   
should be constant in the inner bulge of M31, which is inconsistent with the 
positive radial age gradient. 
Therefore, although we cannot exclude the
possibility that this fly-by interaction produced the young star cluster in the
central 5\arcsec , it cannot explain the intermediate-age stars further beyond. 
Most likely, the intermediate-age stellar population was formed 
in the secular evolution scenario. The presence of a bar in the M31 bulge 
is reported by~\citet{bea07}.~\citet{cou11} also suggest that the ratio of the 
bulge-to-disk scale lengths of M31 is $\sim$0.2, a prediction of secular 
evolution models~\citep{cou96}, which also naturally explains the age gradient and roughly 
constant metallicity of the intermediate-age stellar population. If we assume a 5\% star formation 
efficiency~\citep{eva09} and that disk gas was smoothly transported into the bulge 
during the last 1 Gyr, an infall rate, 3.4 M$_{\odot}$/yr, is expected. This value falls 
within within the range predicted by the numerical 
simulations of the evolution of the galactic disks~\citep[0.1-10 M$_{\odot}$/yr,][]{min12}.

The pseudobulge and classical bulge could co-exist\textbf{~\citep{erw15}}.~\citet{fis10} find that there are a group of pseudobulges 
(defined by their low S\`ersic index and morphology) with very low specific 
star formation rates, just like the classical bulges. Adding a pseudobulge into the 
classical bulge could also efficiently reduce 
its S\`ersic index~\citep{fis10}. This may explain the 
low S\`ersic index of the M31 bulge, compared to the other classical bulges.

Apparently, the suggested secular evolution contributes little to the overall 
mass of the M31 bulge. Assuming secular evolution is responsible for the 
entire intermediate-age stellar population detected, the mass 
fraction of intermediate-age stars derived from our SED fitting
 is indeed small, $\sim$1\%. This is consistent with 
the result in~\citet{mac09}, which suggest that bulge growth via 
secular processes generally contributes little to the stellar mass 
budget in their eight nearby spiral galaxies. Therefore, the M31 bulge is 
similar to the MW bulge, in which the dominant majority of old stars is only 
contaminated by relatively few 
young stars.

\section{Summary}\label{s:summarize}
We have studied stellar populations 
 in the inner bulge of M31 with multi-wavelength observations
taken with \hst\ \wfc3/\acs . The broad wavelength range coverage of our data set 
from near-ultraviolet to near-infrared 
enables us to decompose multiple stellar populations in regions beyond
the central 5\arcsec . We summarize our results below: 

\begin{itemize}
\item In the near-ultraviolet to near-infrared range, the light becomes increasingly 
  blue with the galactocentric distance in the bulge. This trend is in sharp 
  contrast to the positive radial gradient seen in the \galex\ FUV$-$NUV color. 
  The FUV$-$F475W radial color gradient changes at 
  about 50\arcsec\ major-axis radius. These trends cannot be
  explained by the possible presence of old post-EHB stars with a 
negative radial metallicity gradient. 


\item We have found that in addition to the known old stellar population, an
  intermediate-age stellar population is most likely needed to explain the observed
  SED and its radial change across the inner bulge, especially in the three UV bands. 
Although the mass fraction of this new population  
is low (0.2$-$2\%), its emission is significant (e.g., $>$ 40\% in the 
F275W band). 

\item The mass surface density of intermediate-age stars decreases with 
the radius. This trend rules out the possibility that they represent the 
projected galactic disk contribution of M31. We find that this contribution accounts for 
 at most 50\% of the $V$ band intensity in the 
central 100\arcsec\ region. Meanwhile, the metal-poor M31 halo stars can only 
contribute at most extra 3\% the $V$ band intensity in the same region. 

\item The age
  ($\sim$300 Myr$-$1 Gyr) 
and mass fraction of the intermediate-age
  stellar population increase with the galactocentric distance from the center
  of M31, while the metallicity is roughly constant ($\sim$Z$_{\odot}$). 
  
  
 \item Because the intermediate-age stars are older than the young star 
 cluster (age $\sim$200 Myr) in the central 5\arcsec , their formation is 
 unlikely due to the possible head-on collision between M31 and M32. 
 We propose that the population represents the 
  secular evolution of the inner bulge via star formation from 
  the inflow gas, probably induced by the presence of the bar of the galaxy.
  This secular evolution could also explain the positive radial age gradient of the intermediate-age stellar population. 
  
\item The intermediate-age population contributes only $\sim$1\% of the stellar mass in the 
central 180\arcsec , indicating that secular growth plays an insignificant role in building  
the M31 bulge. 

\item The radial age and
  metallicity profiles of the old stellar
  population are consistent with those of~\citet{sag10}; its age
  does not show any significant variation, 
  while its metallicity
  changes from $\sim$ 2Z$_{\odot}$ to $\sim$ Z$_{\odot}$ in the field 
  considered here. This suggests that major mergers taking place in the early 
  Universe produced the majority of old metal-rich stars. Subsequent star formation from the 
  accretion of metal-poor materials may have resulted in the negative metallicity gradient observed. 
\end{itemize}

\section*{Acknowledgments}
\begin{figure*}[!thb]
  \centerline{
      \epsfig{figure=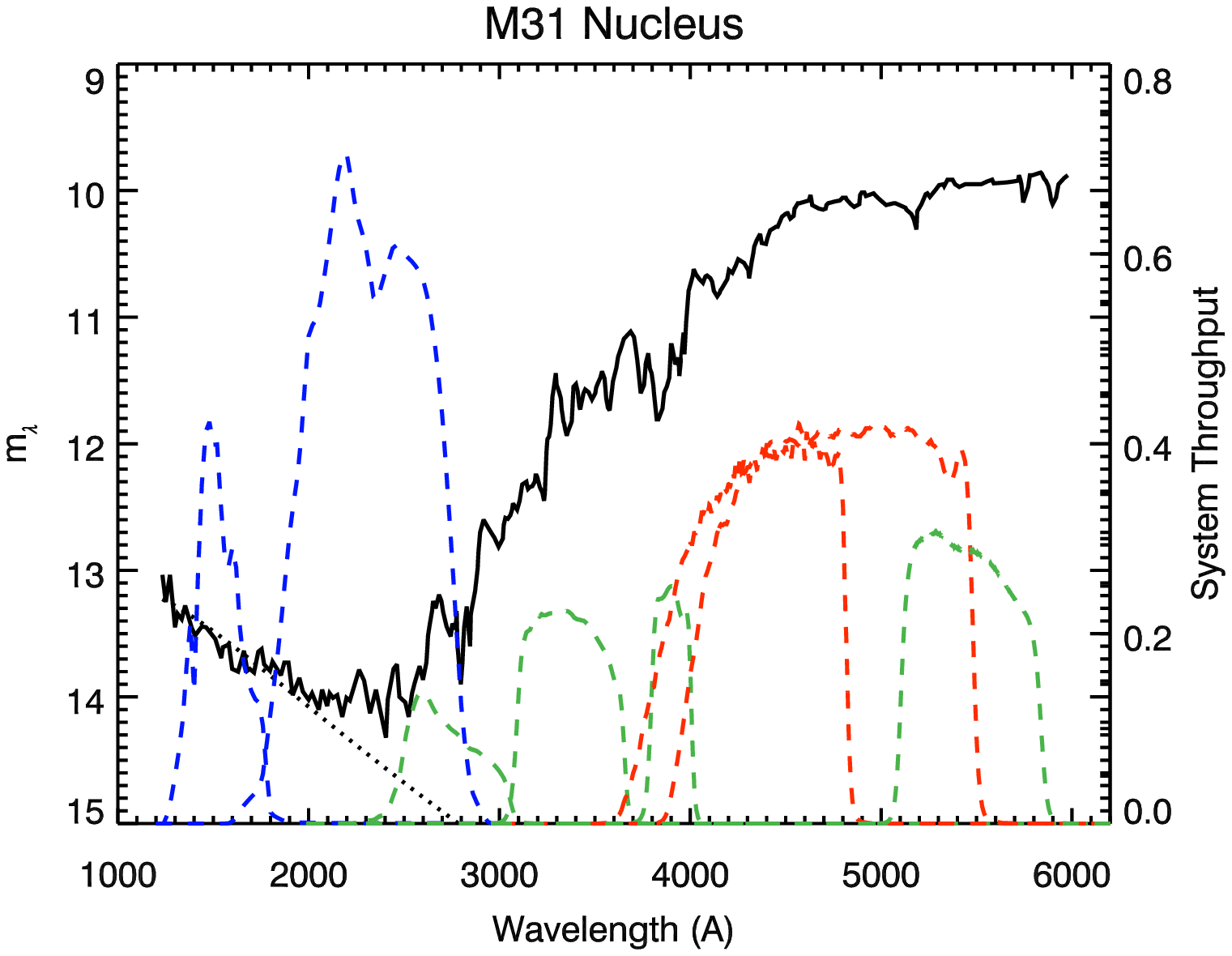,width=1\textwidth,angle=0}
}
 \caption{The spectrum of the central 10\arcsec$\times$20\arcsec\ region 
 of M31 (solid line) obtained from~\citet{oco99}. The units of y-axis are \textbf{Vega} magnitude. 
 The UV part of the spectrum (below 3200 \AA ) is 
 from the International Ultraviolet Explorer, while the part above 3200 \AA\ is from a 
 ground-based telescope. The dotted line represents the EHB contribution 
 inferred from a black-body fit to the spectrum below 2000 \AA. 
 The dashed lines are the transmission curves of the detectors: \galex\ (blue), \hst/WFC3 (green) and 
 \hst/ACS (red) (from left to right, \galex/FUV, \galex/NUV, F275W, F336W, F390M, F435W, 
 F475W, F547M)}. 
 \label{f:ehb}
\end{figure*}

\begin{figure*}[!thb]
  \centerline{
       \epsfig{figure=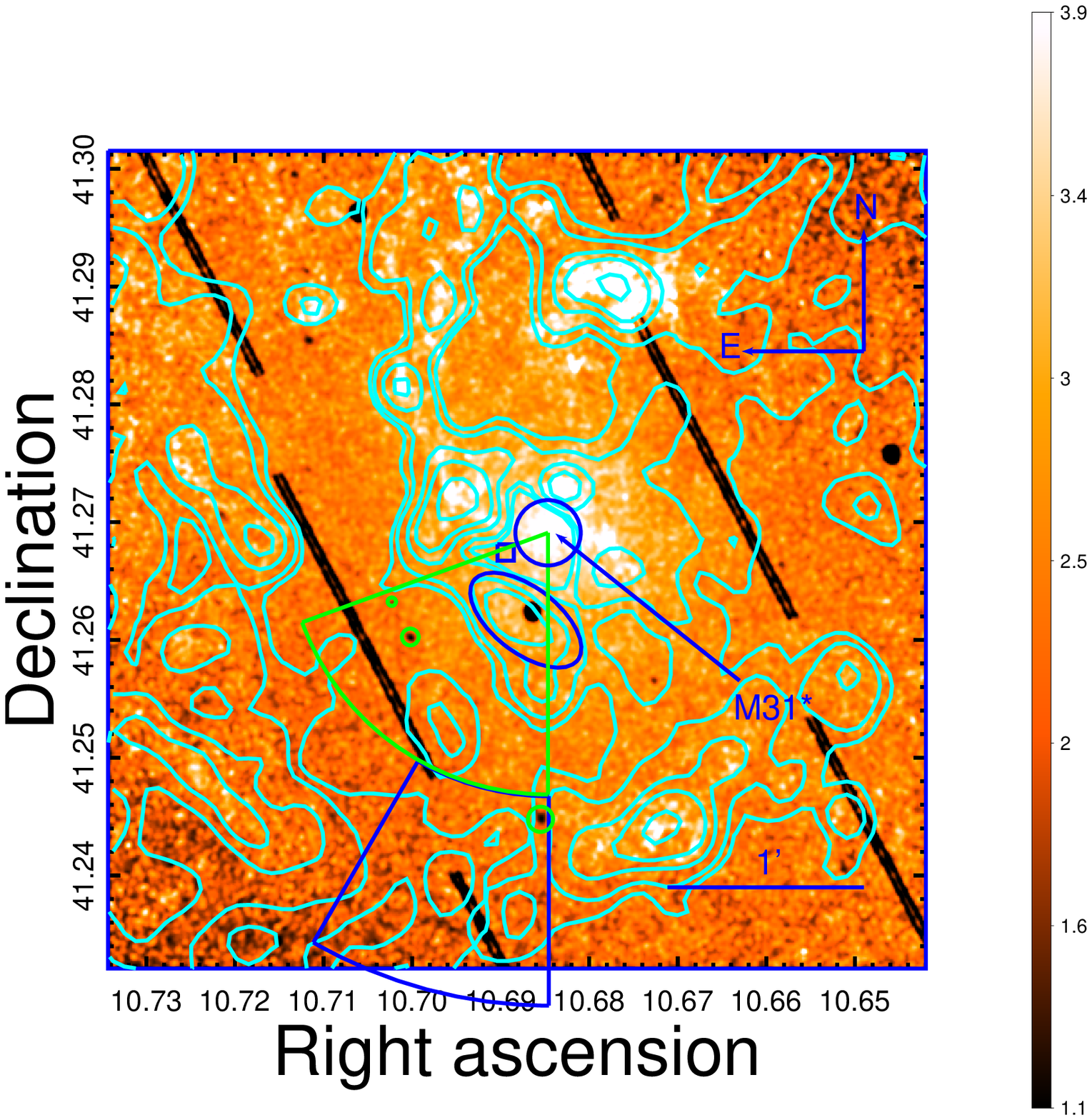,width=1.0\textwidth,angle=0}
       }
 \caption{Map of the F160W to F336W intensity ratio, 
 with overlaid contours from \spitzer/\irac\ 8 $\mu$m 
 dust-only image~\citep{li09}. The blue circle outlines the 
 central 10\arcsec\ region. The
 blue box marks the region used in \S\ref{ss:two} to
 constrain the stellar population in the off-center M31 bulge.
The green and blue sectors are used to study the stellar population variation
  along the minor-axis of the M31 bulge (see \S\ref{ss:under}). Three 
foreground stars and globular clusters (marked with small green circles) are removed. The blue ellipse encompasses 
one dusty clump at 30\arcsec\ ({\it i.e.,} $\sim$120 pc) 
southeast of M31*. The black spot to the southeast is the ‘death star’ (very low
sensitivity) feature in the WFC3 /IR detector. The two black strips are excluded regions,
which were covered by only one dithered exposure in the F275W or F336W bands and do
not allow for cosmic-ray removal.}
 \label{f:ratio}
\end{figure*}

\begin{figure*}[!thb]
  \centerline{
       \epsfig{figure=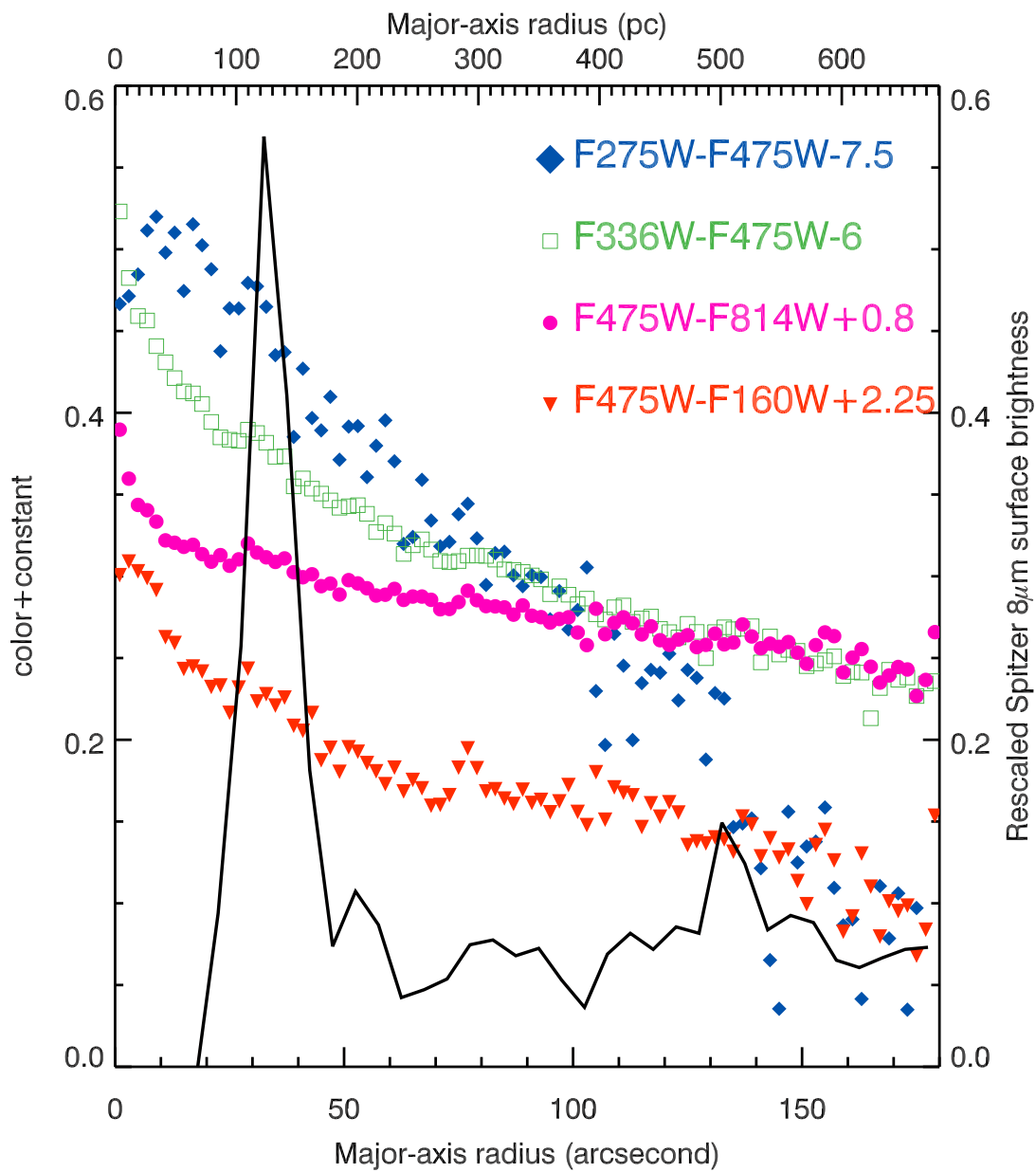,width=1.0\textwidth,angle=0}
       }
 \caption{ The radial distributions of four near-UV to near-IR colors
 along the minor axis of the M31 bulge (see
   \S\ref{ss:under}). The color uncertainties are smaller than the size of the symbols. 
   The distribution of the average surface 
   brightness of \spitzer/\irac\ `dust-only' 8 $\mu$m
   image in the same sky (black curve) is rescaled for comparison.}
\label{f:surface_dis}
 \end{figure*}

\begin{figure*}[!thb]
  \centerline{
      \epsfig{figure=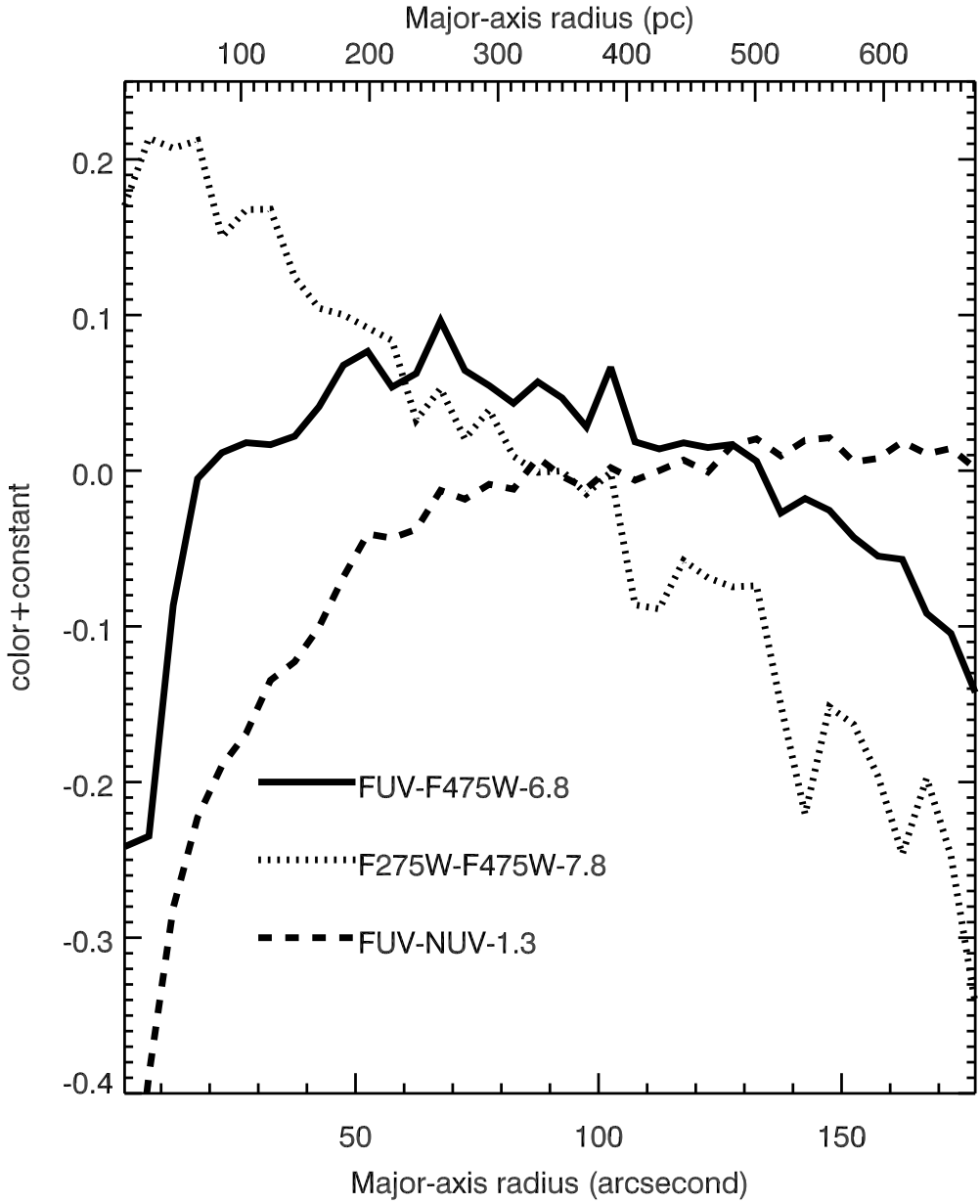,width=1\textwidth,angle=0}
}
 \caption{The radial distributions of the \galex/FUV$-$F475W, F275W$-$F475W
   and \galex\ FUV$-$NUV colors. 
}
\label{f:galex_hst}
\end{figure*}

\begin{figure*}[!thb]
  \centerline{
         \epsfig{figure=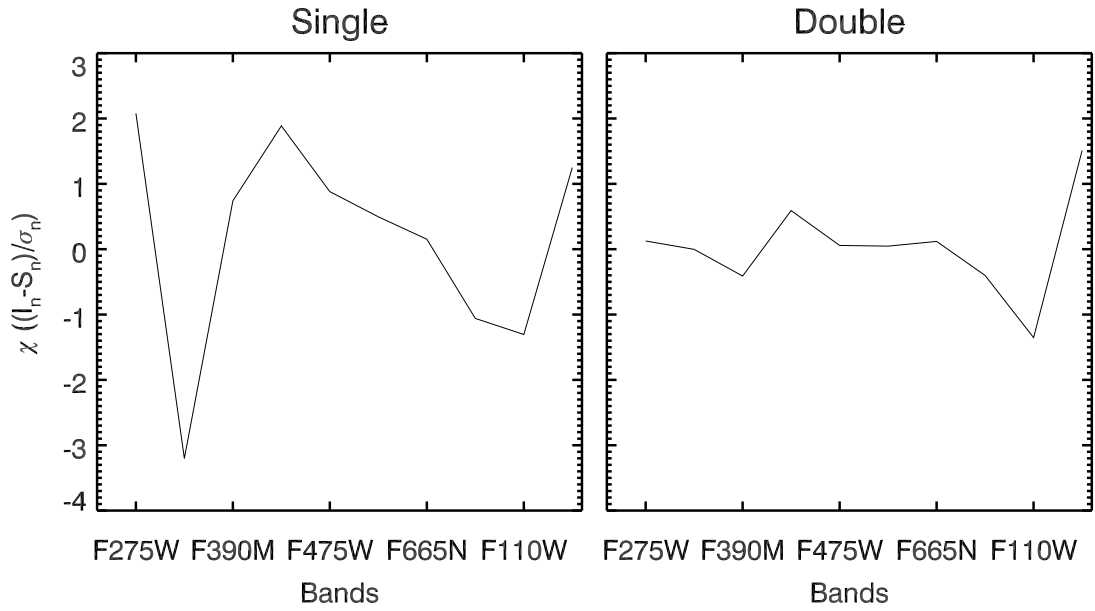,width=1.0\textwidth,angle=0}
      }
 \caption{The $\chi$ values from the modeling with 
   single stellar population (left panel) or intermediate-age+old stellar 
   populations (right panel).}
 \label{f:fit_com}
 \end{figure*}

\begin{figure*}[!thb]
  \centerline{
        \epsfig{figure=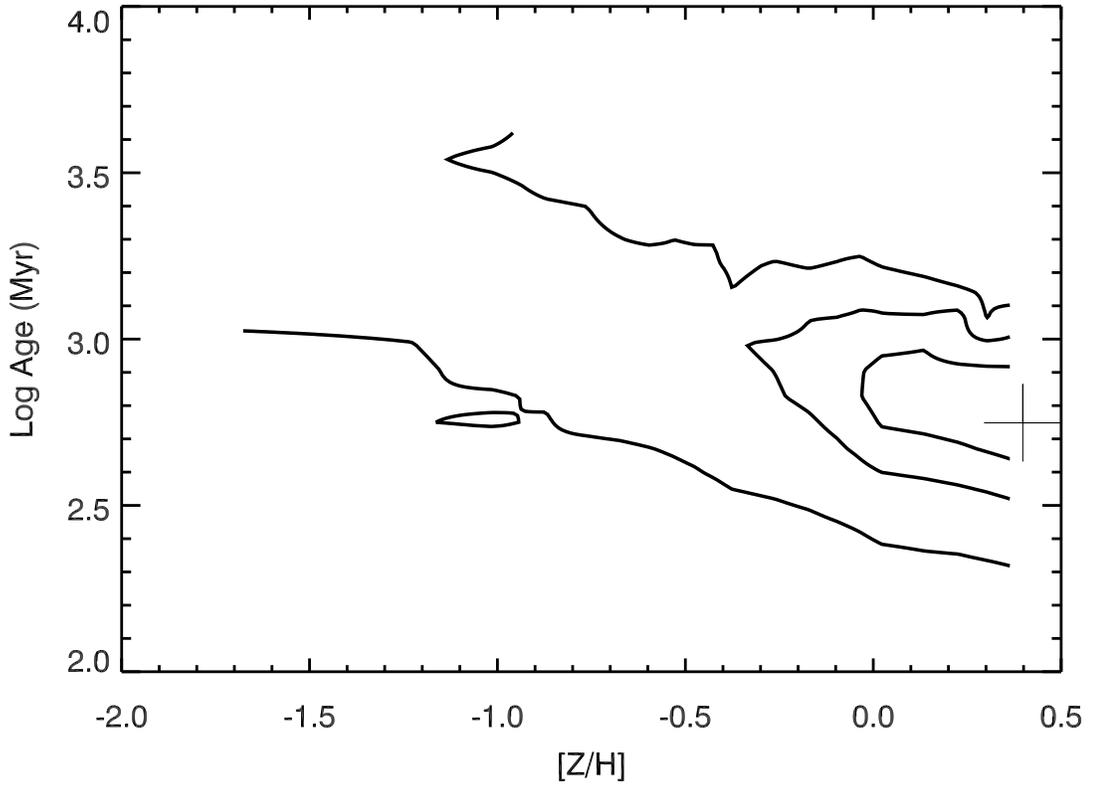,width=1.0\textwidth,angle=0}
      }
      \caption{The relationship between the metallicity
        ([Z/H]=log$_{10}$(Z/Z$_{\odot}$)) and age of the
  intermediage-age stellar population, as inferred from 
  the two-population SED fitting for a randomly
  selected region in the M31 bulge (blue box in Fig.~\ref{f:ratio}). 
  The contours are at the 
  68, 90 and 95\% confidence levels. The `plus' symbol marks the 
best-fit metallicity and age. \textbf{Limited by the stellar 
synthesis model, our contours are not extended to [Z/H] $>$ 0.4.}}
\label{f:ya_ym}
 \end{figure*}

\begin{figure*}[!thb]
  \centerline{
      \epsfig{figure=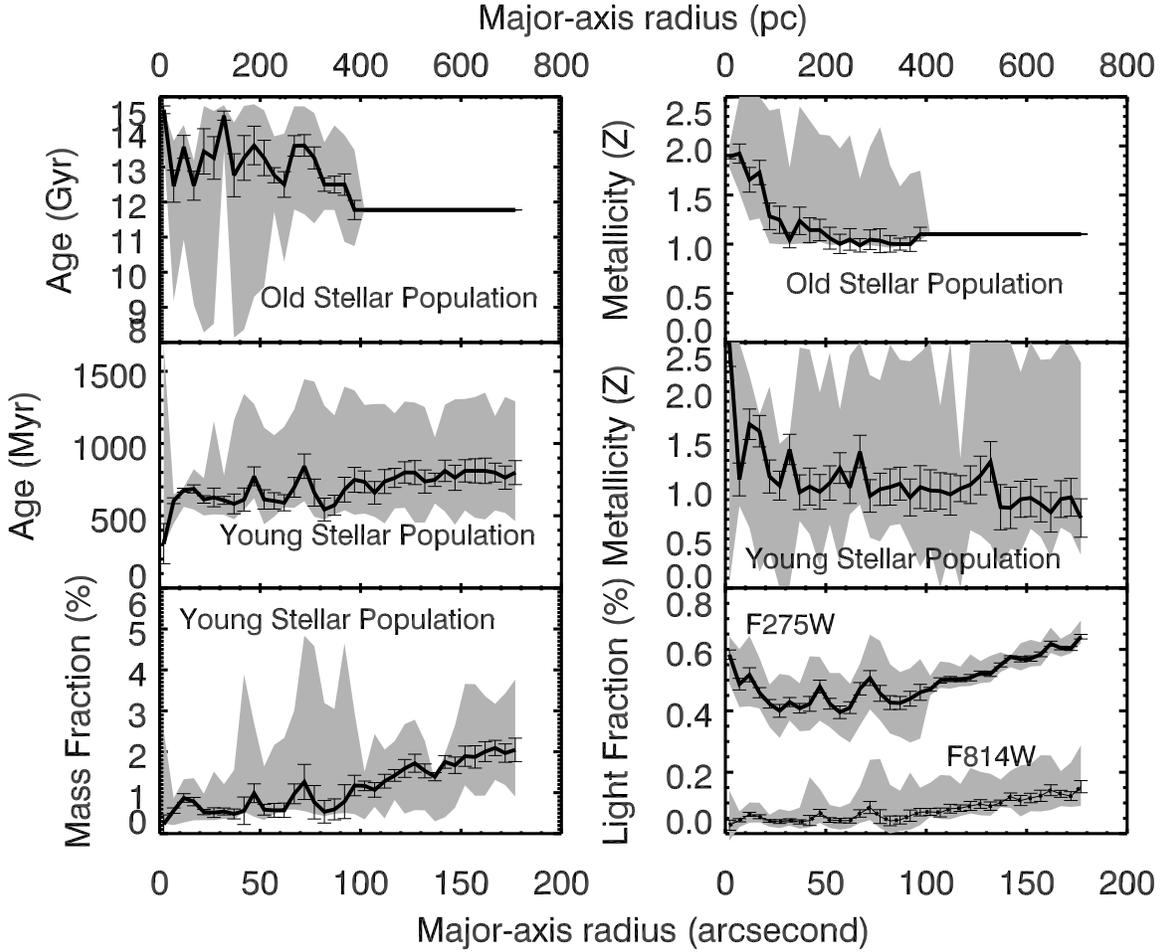,width=1.1\textwidth,angle=0}
}
 \caption{The radial distributions of the stellar age, metallicity,  
 mass and light fraction for the young (intermediate-age) and old 
 stellar population, as marked in individual panels. The bottom right panel shows the light 
 fraction for the young (intermediate-age) stellar population in the integrated light in the F275W 
 and F814W bands. 
 In each panel, the
 grey area represents the uncertainty range at the 68\% confidence level, while the error bars are
 the uncertainties of the means at individual radial intervals. \textbf{Due to the limit of data points (see Section~\ref{ss:under}), 
 beyond 100\arcsec ,  we fix the age and metallicity of the old stellar population to the corresponding values in the annulus of 
 95\arcsec\ to 100\arcsec .}}
 \label{f:annu_chi}
 \end{figure*}

\begin{figure*}[!thb]
  \centerline{
      \epsfig{figure=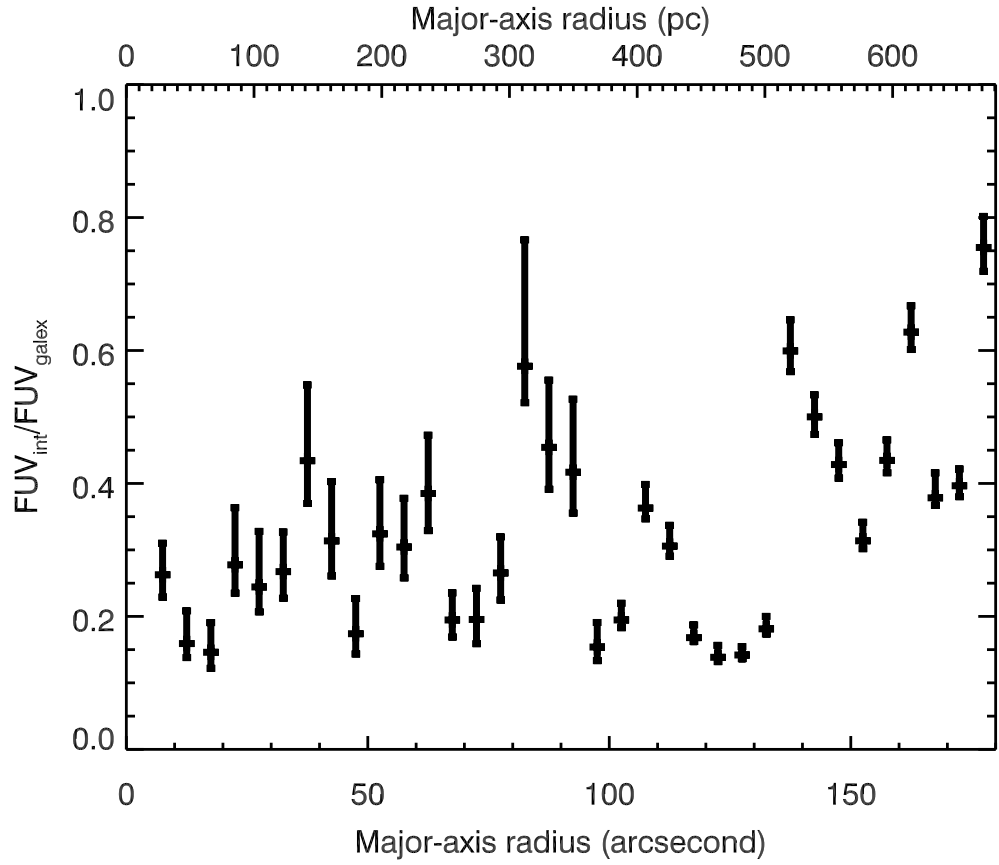,width=1\textwidth,angle=0}
}
 \caption{The radial distribution of the light contribution for the intermediate-age
  stellar population (FUV$_{int}$) in the observed 
 \galex\ FUV band (FUV$_{galex}$,~\citealt{thi05}).} 
 \label{f:fuv_int_galex}
\end{figure*}

\begin{figure*}[!thb]
  \centerline{
      \epsfig{figure=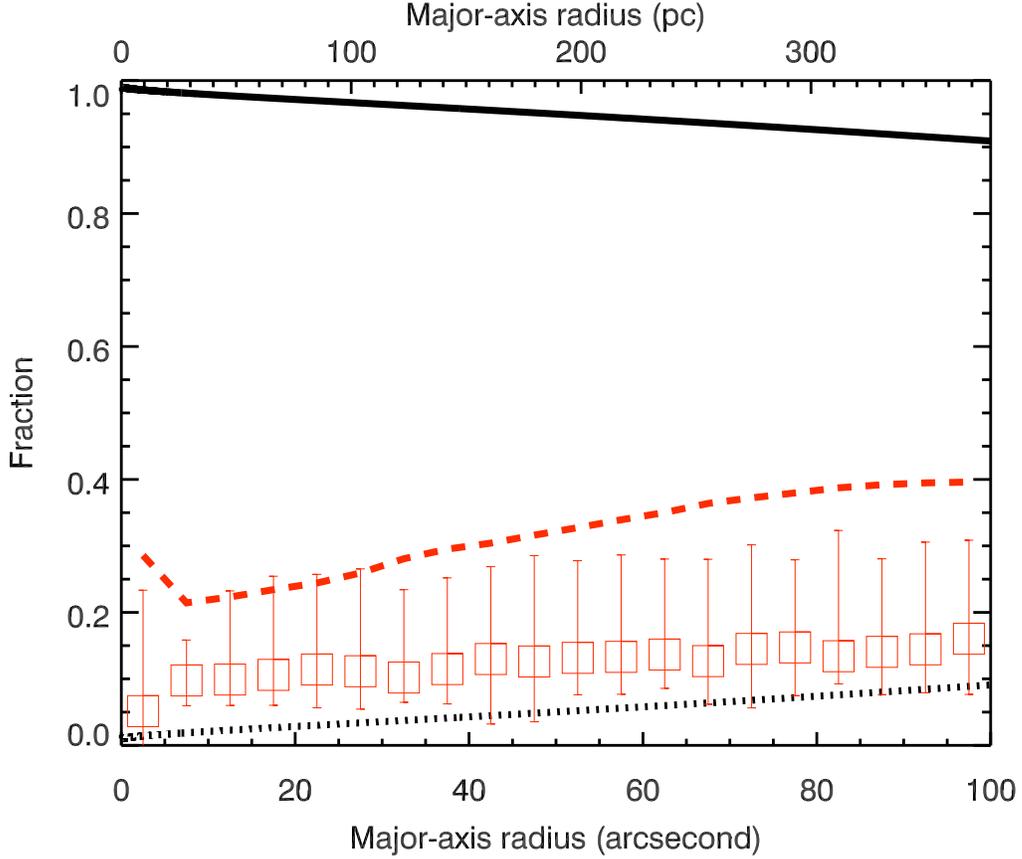,width=1.0\textwidth,angle=0}
}
 \caption{Comparison between various V-band/F547M intensity fractions. 
 The black solid and dotted lines represent the total 
 fractions contributed by the bulge and disk of M31 in the $V$-band, respectively~\citep{kor99}. 
 The square and error signs represent the fraction from the intermediate-age stellar
 population and its 68\% uncertainty in the observed F547M intensity as estimated 
 from our SED fitting. Because there is no F547M data beyond 100\arcsec , we  
 show only the curves within this region. The red dashed line is for the fractional contribution 
 of the disk to the intermediate-age stellar population ({\it i.e.}, the squares divided by the black dotted line).}
 \label{f:disk_con}
 \end{figure*}

\begin{figure*}[!thb]
  \centerline{
      \epsfig{figure=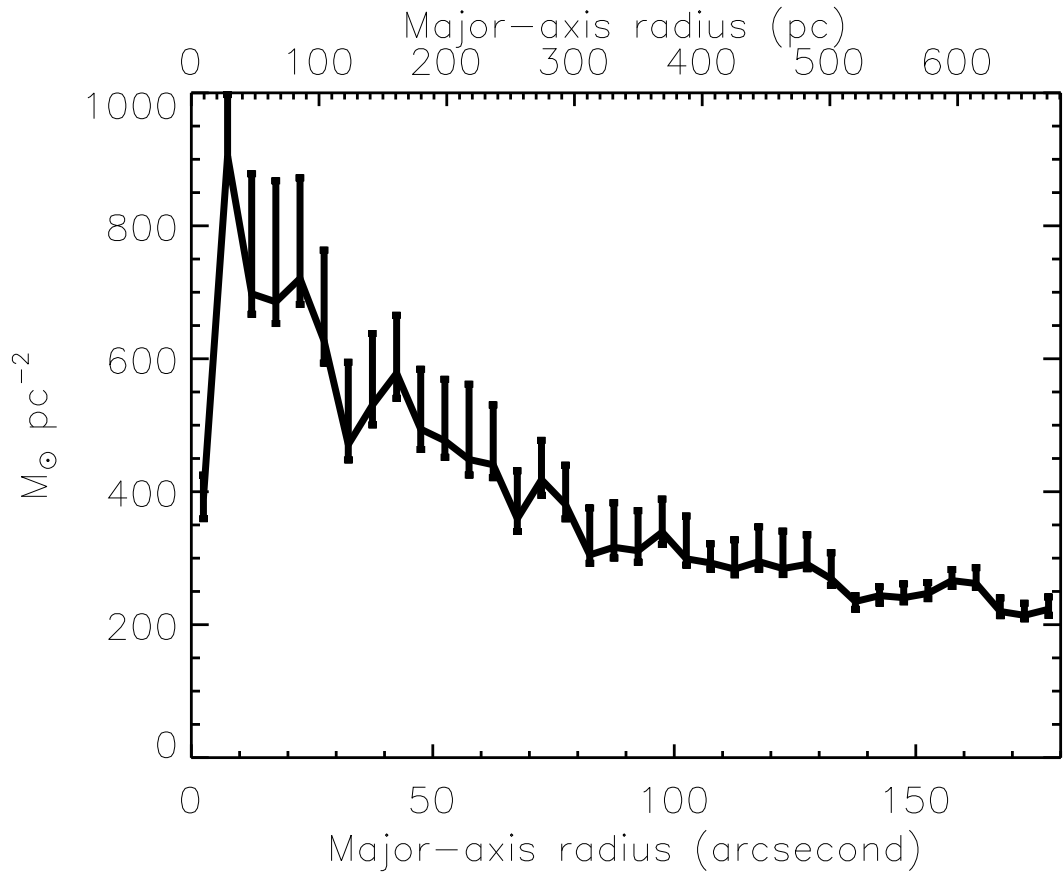,width=1.0\textwidth,angle=0}
}
 \caption{The mass surface density of the 
 intermediate-age stellar population as a function of the major-axis radius.}
  \label{f:disk_mass}
 \end{figure*}

This article is based on observations made 
with the NASA/ESA Hubble Space Telescope and obtained 
from the data archive at the Space Telescope Science Institute, which is 
operated by the Association of Universities for Research in Astronomy, 
Inc. under NASA contract NAS 5-26555. We are grateful to Philip Rosenfield, 
Luciana Bianchi, 
Antonela Monachesi, Morgan Fouesneau for valuable
comments and discussion. This work is supported by NASA grant GO-12055 
 provided by the Space Telescope Science Institute, which
is operated by the Association of Universities for Research in
Astronomy, Inc., under NASA contract NAS 5-26555. H. D. 
acknowledges the support and hospitality of the Key Laboratory 
of Modern Astronomy and Astrophysics at 
Nanjing University during his visit, and would like to thank Robert W. O'Connell 
and Daniela Calzetti for providing the the spectrum used in Figure~\ref{f:ehb}. Z.L. acknowledges 
support from the Recruitment Program of Global Youth Experts and the National Natural Science 
Foundation of China (grant 11133001).

\begin{deluxetable}{ccccccccccc}
\rotate
  \tabletypesize{\tiny}
 \tablecolumns{11}
  \tablecaption{\hst\ Multi-Wavelength Observations}
  \tablewidth{0pt}
  \tablehead{
  \colhead{}&
  \colhead{}&
  \colhead{Pivot $\lambda$} &
   \colhead{Program} & 
  \colhead{Number of} &
  \colhead{Exposure} &
  \colhead{PHOTFLAM} &
   \colhead{Systematic error} & 
   \colhead{Median of } & 
   \colhead{$A_n/A_{F547M}$} &  
   \colhead{$A_n/A_{F547M}$}\\
  \colhead{Filter} & 
  \colhead{Detector} &
  \colhead{($\dot{A}$)} &
  \colhead{ID} &
  \colhead{dithered exposures} &
  \colhead{time (s)}  &
  \colhead{ergs $cm^{-2}$ $s^{-1}$ $\dot{A}^{-1}$} &
  \colhead{of PHOTFLAM$^{a}$} & 
  \colhead{$\sigma$/I (\%)$^{b}$} & 
  \colhead{Milky Way$^c$} & 
  \colhead{M31 bulge$^d$} \\
  }
  \startdata
F275W & \wfc3 /UVIS & 2704 & 12058&2& 925 & 3.3010e-18 & 3.5\% &
13.1&1.94 & 2.80\\
F336W & \wfc3 /UVIS & 3355 & 12058&2& 1250 & 1.3129e-18 & 2\% & 3.7 &
1.64 & 1.92\\
F390M & \wfc3 /UVIS & 3897 & 12174 & 3 & 2700 & 2.5171e-18 & 2\% & 3.1
& 1.47 & 1.58\\
F435W & \acs /WFC & 4319 & 10006 & 4 & 2200 & & & & &\\
F435W & \acs /WFC & 4319 & 10760 & 8 & 4360 & 3.1840e-19 & 2\% &
3.6 & 1.31 & 1.37\\
F435W & \acs /WFC & 4319 & 11833 &  8 & 4360 & & & & &\\
F475W & \acs /WFC & 4747 & 12058 & 5 & 1900 &1.8210e-19 & 2\% & 4.1 &
1.18 & 1.21\\
F547M & \wfc3 /UVIS &  5447 & 12174 & 3 & 2700 & 4.6321e-19 & 2\% &
4.1 & 1.0 & 1.0\\
F665N & \wfc3 /UVIS &  6656 & 12174 & 3 & 2700 & 1.9943e-18 & 2\% &
5.0 & 0.75 & 0.73\\ 
F814W & \acs /WFC & 8057 & 12058 & 4 & 1715 & 7.0332e-20 & 2\% & 7.8 &
0.55 & 0.61\\
F110W & \wfc3 /IR & 11534 & 12058 & 1 & 699 & 1.5274e-20 & 2\% & 12.7 &
0.31 & 0.46\\
F160W & \wfc3 /IR &  15369 & 12058 & 4& 1600 & 1.9276e-20 & 2\% & 14.7
& 0.19 & 0.43\\
 \enddata
\label{t:obs}
\tablecomments{ a) The `PHOTFLAM' information are from:
  http://www.stsci.edu/hst/wfc3/phot\_zp\_lbn (WFC3) and
  http://www.stsci.edu/hst/acs/analysis/zeropoints/\#tablestart (ACS).
b) $\sigma$ and `I' are the uncertainty and intensity of
  each pixel. c) The relative extinction
  $A_n/A_{F547M}$ for the MW-type dust. d) The average
  relative extinction $A_n/A_{F547M}$ of five dusty clumps in the M31 bulge, derived
  by~\citet{don14}. }
\end{deluxetable}

\begin{deluxetable}{ccccccccc}
  \tabletypesize{\scriptsize}
 \tablecolumns{10}
  \tablecaption{SED fitting for the southwest region of the M31 bulge}
  \tablewidth{0pt}
  \tablehead{
  \colhead{Major-axis}&
  \colhead{$t_{old}$} &
  \colhead{$Z_{old}$} & 
  \colhead{$M_{old}$} &
  \colhead{$t_{new}$} &
  \colhead{$Z_{new}$} &
  \colhead{$M_{new}$} &
   \colhead{Mass fraction of the } & 
  \colhead{Reduced }\\
  \colhead{radius(\arcsec )}&
  \colhead{(Gyr)} & 
  \colhead{($Z_{\odot}$)} &
  \colhead{($10^6M_{\odot}$)} &
  \colhead{(Myr)} & 
  \colhead{($Z_{\odot}$)} &
  \colhead{($10^6M_{\odot}$)} &
  \colhead{young population (\%)} &
  \colhead{$\chi^2$}\\
}
  \startdata
0-5&$14.6_{-1.1}^{+0.0}$&$1.89_{-0.11}^{+0.11}$&$ 31.1_{-0.1}^{+4.0}$&$307_{-14}^{+1374}$&$2.50_{-2.45}^{+0.00}$&$ 0.1_{-0.0}^{+0.1}$&$0.2_{-0.0}^{+2.3}$&2.74\\
5-10&$12.5_{-3.3}^{+1.3}$&$1.92_{-0.40}^{+0.58}$&$ 53.4_{-0.4}^{+0.7}$&$604_{-157}^{+42}$&$1.10_{-0.27}^{+1.40}$&$ 0.3_{-0.1}^{+0.3}$&$0.6_{-0.4}^{+0.3}$&1.53\\
10-15&$13.6_{-2.6}^{+0.6}$&$1.66_{-0.41}^{+0.84}$&$ 70.1_{-1.0}^{+1.0}$&$677_{-115}^{+21}$&$1.67_{-1.05}^{+0.50}$&$ 0.6_{-0.3}^{+0.6}$&$0.9_{-0.6}^{+0.5}$&1.38\\
15-20&$12.5_{-3.4}^{+0.7}$&$1.73_{-0.47}^{+0.77}$&$ 83.3_{-0.7}^{+1.8}$&$686_{-142}^{+136}$&$1.60_{-1.43}^{+0.17}$&$ 0.7_{-0.3}^{+0.6}$&$0.8_{-0.5}^{+0.6}$&1.36\\
20-25&$13.4_{-5.2}^{+1.3}$&$1.28_{-0.28}^{+1.08}$&$ 97.9_{-0.3}^{+1.3}$&$613_{-110}^{+147}$&$1.14_{-0.76}^{+0.92}$&$ 0.5_{-0.3}^{+0.5}$&$0.5_{-0.2}^{+0.4}$&1.33\\
25-30&$13.3_{-4.7}^{+1.3}$&$1.25_{-0.25}^{+1.13}$&$ 89.9_{-0.0}^{+2.8}$&$627_{-117}^{+528}$&$1.04_{-1.04}^{+0.43}$&$ 0.5_{-0.2}^{+0.4}$&$0.5_{-0.2}^{+1.0}$&1.52\\
30-35&$14.5_{-1.0}^{+0.3}$&$1.04_{-0.04}^{+0.74}$&$104.5_{-0.1}^{+2.6}$&$607_{-100}^{+173}$&$1.41_{-1.39}^{+0.15}$&$ 0.6_{-0.3}^{+0.5}$&$0.5_{-0.2}^{+0.7}$&1.68\\
35-40&$12.8_{-4.6}^{+1.5}$&$1.24_{-0.23}^{+1.16}$&$111.2_{-0.1}^{+4.3}$&$582_{-107}^{+578}$&$0.98_{-0.29}^{+1.47}$&$ 0.5_{-0.3}^{+0.5}$&$0.5_{-0.1}^{+1.1}$&1.25\\
40-45&$13.3_{-4.9}^{+1.5}$&$1.14_{-0.14}^{+1.12}$&$116.6_{-0.7}^{+12.4}$&$612_{-71}^{+731}$&$1.03_{-0.49}^{+1.39}$&$ 0.7_{-0.3}^{+0.5}$&$0.6_{-0.0}^{+3.3}$&1.22\\
45-50&$13.6_{-4.3}^{+1.2}$&$1.14_{-0.12}^{+0.97}$&$115.1_{-1.3}^{+5.1}$&$773_{-150}^{+500}$&$0.98_{-0.76}^{+1.01}$&$ 1.1_{-0.6}^{+1.0}$&$1.0_{-0.6}^{+1.7}$&1.17\\
50-55&$13.3_{-3.7}^{+1.4}$&$1.06_{-0.07}^{+1.04}$&$119.7_{-0.7}^{+4.8}$&$613_{-171}^{+496}$&$1.07_{-0.42}^{+1.33}$&$ 0.7_{-0.3}^{+0.6}$&$0.6_{-0.3}^{+1.1}$&1.13\\
55-60&$12.8_{-1.0}^{+1.1}$&$1.00_{-0.01}^{+0.97}$&$117.4_{-0.2}^{+4.9}$&$603_{-109}^{+454}$&$1.22_{-0.97}^{+0.60}$&$ 0.7_{-0.3}^{+0.6}$&$0.6_{-0.1}^{+1.6}$&1.16\\
60-65&$12.5_{-2.2}^{+1.4}$&$1.05_{-0.04}^{+1.06}$&$114.8_{-0.8}^{+6.6}$&$590_{-30}^{+525}$&$1.02_{-0.27}^{+1.34}$&$ 0.7_{-0.3}^{+0.6}$&$0.6_{-0.0}^{+1.8}$&1.30\\
65-70&$13.6_{-1.8}^{+1.1}$&$0.99_{-0.01}^{+0.68}$&$114.3_{-0.5}^{+6.2}$&$697_{-142}^{+540}$&$1.39_{-0.54}^{+1.11}$&$ 1.1_{-0.6}^{+1.0}$&$1.0_{-0.4}^{+2.0}$&1.38\\
70-75&$13.6_{-2.0}^{+1.1}$&$1.05_{-0.06}^{+0.95}$&$114.9_{-1.8}^{+12.3}$&$843_{-241}^{+602}$&$0.94_{-0.43}^{+1.23}$&$ 1.5_{-0.7}^{+1.1}$&$1.3_{-0.7}^{+3.6}$&1.47\\
75-80&$13.3_{-1.9}^{+1.3}$&$1.04_{-0.06}^{+1.16}$&$105.4_{-0.3}^{+14.3}$&$660_{-166}^{+768}$&$1.00_{-0.53}^{+1.11}$&$ 0.8_{-0.4}^{+0.6}$&$0.8_{-0.3}^{+3.8}$&1.43\\
80-85&$12.5_{-0.7}^{+1.2}$&$1.00_{-0.04}^{+0.80}$&$ 99.0_{-0.2}^{+11.2}$&$543_{-68}^{+718}$&$1.03_{-0.39}^{+1.45}$&$ 0.5_{-0.3}^{+0.4}$&$0.5_{-0.1}^{+2.7}$&1.63\\
85-90&$12.5_{-0.7}^{+1.7}$&$1.00_{-0.02}^{+0.59}$&$104.0_{-0.3}^{+8.7}$&$573_{-65}^{+622}$&$1.06_{-0.41}^{+1.26}$&$ 0.6_{-0.3}^{+0.5}$&$0.6_{-0.1}^{+2.5}$&1.72\\
90-95&$12.5_{-1.6}^{+1.4}$&$1.00_{-0.02}^{+0.72}$&$102.4_{-0.7}^{+11.7}$&$681_{-142}^{+713}$&$0.92_{-0.49}^{+1.39}$&$ 0.8_{-0.4}^{+0.6}$&$0.8_{-0.3}^{+3.9}$&1.82\\
95-100&$11.8_{-1.0}^{+1.7}$&$1.10_{-0.06}^{+0.65}$&$ 86.8_{-0.6}^{+5.4}$&$749_{-241}^{+618}$&$1.03_{-0.81}^{+1.37}$&$ 1.0_{-0.5}^{+0.9}$&$1.2_{-0.7}^{+1.8}$&1.97\\
100-105&11.8&1.10&$ 33.8_{-0.5}^{+1.9}$&$735_{-219}^{+532}$&$0.99_{-0.60}^{+1.51}$&$ 0.4_{-0.2}^{+0.3}$&$1.2_{-0.8}^{+0.3}$&1.38\\
105-110&11.8&1.10&$ 42.7_{-0.1}^{+4.2}$&$659_{-165}^{+615}$&$0.99_{-0.99}^{+0.87}$&$ 0.5_{-0.2}^{+0.4}$&$1.1_{-0.4}^{+1.4}$&1.52\\
110-115&11.8&1.10&$ 42.4_{-0.8}^{+3.6}$&$737_{-289}^{+536}$&$0.95_{-0.55}^{+1.55}$&$ 0.5_{-0.3}^{+0.5}$&$1.3_{-0.9}^{+0.8}$&1.45\\
115-120&11.8&1.10&$ 41.7_{-0.1}^{+2.5}$&$762_{-211}^{+446}$&$1.01_{-1.01}^{+0.31}$&$ 0.6_{-0.3}^{+0.5}$&$1.4_{-0.6}^{+0.8}$&1.51\\
120-125&11.8&1.10&$ 41.1_{-0.7}^{+3.5}$&$800_{-244}^{+490}$&$1.05_{-0.46}^{+1.45}$&$ 0.7_{-0.3}^{+0.5}$&$1.6_{-1.1}^{+1.1}$&1.68\\
125-130&11.8&1.10&$ 39.8_{-0.7}^{+2.9}$&$799_{-345}^{+484}$&$1.15_{-0.61}^{+1.35}$&$ 0.7_{-0.3}^{+0.6}$&$1.7_{-1.1}^{+1.1}$&1.45\\
130-135&11.8&1.10&$ 36.9_{-0.5}^{+3.0}$&$737_{-284}^{+515}$&$1.28_{-0.84}^{+1.22}$&$ 0.6_{-0.3}^{+0.5}$&$1.5_{-1.0}^{+0.7}$&1.37\\
135-140&11.8&1.10&$ 36.1_{-1.4}^{+0.7}$&$749_{-164}^{+294}$&$0.82_{-0.62}^{+1.68}$&$ 0.5_{-0.2}^{+0.5}$&$1.4_{-1.1}^{+0.0}$&1.30\\
140-145&11.8&1.10&$ 34.8_{-1.7}^{+1.2}$&$811_{-281}^{+458}$&$0.82_{-0.42}^{+1.68}$&$ 0.6_{-0.3}^{+0.6}$&$1.8_{-1.4}^{+0.2}$&1.27\\
145-150&11.8&1.10&$ 34.7_{-0.9}^{+3.2}$&$763_{-325}^{+543}$&$0.90_{-0.54}^{+1.30}$&$ 0.6_{-0.3}^{+0.5}$&$1.7_{-1.0}^{+1.0}$&1.19\\
150-155&11.8&1.10&$ 33.9_{-0.3}^{+3.2}$&$811_{-274}^{+497}$&$0.92_{-0.32}^{+1.58}$&$ 0.7_{-0.3}^{+0.5}$&$1.9_{-0.7}^{+1.8}$&1.14\\
155-160&11.8&1.10&$ 27.6_{-0.8}^{+2.7}$&$811_{-324}^{+510}$&$0.85_{-0.33}^{+1.47}$&$ 0.5_{-0.3}^{+0.4}$&$1.9_{-0.9}^{+1.8}$&1.03\\
160-165&11.8&1.10&$ 21.0_{-0.9}^{+2.0}$&$811_{-336}^{+541}$&$0.77_{-0.45}^{+1.56}$&$ 0.4_{-0.2}^{+0.3}$&$2.0_{-1.1}^{+1.4}$&0.98\\
165-170&11.8&1.10&$ 15.5_{-0.2}^{+1.0}$&$799_{-261}^{+391}$&$0.91_{-0.29}^{+1.58}$&$ 0.3_{-0.2}^{+0.3}$&$2.1_{-0.8}^{+1.1}$&0.97\\
170-175&11.8&1.10&$ 11.3_{-0.1}^{+1.0}$&$764_{-243}^{+558}$&$0.92_{-0.37}^{+1.55}$&$ 0.2_{-0.1}^{+0.2}$&$2.0_{-0.8}^{+1.5}$&0.94\\
175-180&11.8&1.10&$  7.4_{-0.4}^{+0.7}$&$799_{-336}^{+492}$&$0.71_{-0.38}^{+1.58}$&$ 0.2_{-0.1}^{+0.1}$&$2.0_{-1.1}^{+1.7}$&0.92\\

\enddata
\label{t:annu_fit}
\tablecomments{The superscript and subscript represent the 68\% 
confidence range. Beyond the 100\arcsec , there are no
  F390M, F547M and F665N observations. In order to have enough freedom for the
least $\chi^2$ fitting, we freeze the
  age and metallicity of the old stellar population to the values at
  the annulus 95\arcsec -100\arcsec .} 
\end{deluxetable}


\begin{thebibliography}{}
\bibitem[Athanassoula(2005)]{ath05} Athanassoula, E.\ 2005, 
\mnras, 358, 1477 
\bibitem[Beaton et al.(2007)]{bea07} Beaton, R.~L., Majewski, 
S.~R., Guhathakurta, P., et al.\ 2007, \apjl, 658, L91 
\bibitem[Bender et al.(2005)]{ben05} Bender, R., Kormendy, 
J., Bower, G., et al.\ 2005, \apj, 631, 280 
\bibitem[Bensby et 
al.(2013)]{ben13} Bensby, T., Yee, J.~C., Feltzing, S., et al.\ 2013, \aap, 549, A147 
\bibitem[Block et al.(2006)]{blo06} Block, D.~L., Bournaud, 
F., Combes, F., et al.\ 2006, \nat, 443, 832
\bibitem[Bressan et al.(2012)]{bre12} Bressan, A., Marigo, 
P., Girardi, L., et al.\ 2012, \mnras, 427, 127 
\bibitem[Brown et al.(1998)]{bro98} Brown, T.~M., Ferguson, 
H.~C., Stanford, S.~A., \& Deharveng, J.-M.\ 1998, \apj, 504, 113 
\bibitem[Bureau 
\& Athanassoula(2005)]{bur05} Bureau, M., \& Athanassoula, E.\ 2005, \apj, 626, 159 
\bibitem[Crane et al.(1992)]{cra92} Crane, P.~C., Dickel, 
J.~R., \& Cowan, J.~J.\ 1992, \apjl, 390, L9
\bibitem[Conroy 
\& van Dokkum(2012)]{con12} Conroy, C., \& van Dokkum, P.~G.\ 2012, \apj, 760, 71 
\bibitem[Courteau(1996)]{cou96} Courteau, S.\ 1996, \apjs, 
103, 363 
\bibitem[Courteau et al.(2011)]{cou11} Courteau, S., Widrow, 
L.~M., McDonald, M., et al.\ 2011, \apj, 739, 20
\bibitem[Dalcanton et al.(2012)]{dal12} Dalcanton, J.~J., 
Williams, B.~F., Lang, D., et al.\ 2012, \apjs, 200, 18 
\bibitem[Dierickx et al.(2014)]{die14} Dierickx, M., Blecha, 
L., \& Loeb, A.\ 2014, \apjl, 788, L38 
\bibitem[Dong et al.(2011)]{don11} Dong, H., Wang, Q.~D., 
Cotera, A., et al.\ 2011, \mnras, 417, 114 
\bibitem[Dong et al.(2014)]{don14} Dong, H., Li, Z., Wang, 
Q.~D., et al.\ 2014, \apj, 785, 136 
\bibitem[Dorman et al.(2013)]{dor13} Dorman, C.~E., Widrow, 
L.~M., Guhathakurta, P., et al.\ 2013, \apj, 779, 103
\bibitem[Dressler 
\& Richstone(1988)]{dre88} Dressler, A., \& Richstone, D.~O.\ 1988, \apj, 324, 701
\bibitem[Erwin et al.(2015)]{erw15} Erwin, P., Saglia, R.~P., 
Fabricius, M., et al.\ 2015, \mnras, 446, 4039 
\bibitem[Evans et al.(2009)]{eva09} Evans, N.~J., II, Dunham, 
M.~M., J{\o}rgensen, J.~K., et al.\ 2009, \apjs, 181, 321 
\bibitem[Fisher 
\& Drory(2010)]{fis10} Fisher, D.~B., \& Drory, N.\ 2010, \apj, 716, 942 
\bibitem[Fitzpatrick(1999)]{fit99} Fitzpatrick, E.~L.\ 1999, 
\pasp, 111, 63 
\bibitem[Garcia et al.(2010)]{gar10} Garcia, M.~R., Hextall, 
R., Baganoff, F.~K., et al.\ 2010, \apj, 710, 755 
\bibitem[Ibata et al.(2014)]{iba14} Ibata, R.~A., Lewis, 
G.~F., McConnachie, A.~W., et al.\ 2014, \apj, 780, 128 
\bibitem[Kalirai et al.(2006)]{kal06} Kalirai, J.~S., 
Gilbert, K.~M., Guhathakurta, P., et al.\ 2006, \apj, 648, 389 
\bibitem[Kaviraj et al.(2007)]{kav07} Kaviraj, S., Rey, 
S.-C., Rich, R.~M., Yoon, S.-J., \& Yi, S.~K.\ 2007, \mnras, 381, L74 
\bibitem[King et al.(1995)]{kin95} King, I.~R., Stanford, 
S.~A., \& Crane, P.\ 1995, \aj, 109, 164 
\bibitem[Kormendy(1988)]{kor88} Kormendy, J.\ 1988, \apj, 
325, 128
\bibitem[Kormendy 
\& Bender(1999)]{kor99} Kormendy, J., \& Bender, R.\ 1999, \apj, 522,
772 
\bibitem[Kormendy 
\& Kennicutt(2004)]{kor04} Kormendy, J., \& Kennicutt, R.~C., Jr.\ 2004, \araa, 42, 603 
\bibitem[Kroupa(2002)]{kro02} Kroupa, P.\ 2002, Modes of Star 
Formation and the Origin of Field Populations, 285, 86 
\bibitem[Lauer et al.(1998)]{lau98} Lauer, T.~R., Faber, 
S.~M., Ajhar, E.~A., Grillmair, C.~J., 
\& Scowen, P.~A.\ 1998, \aj, 116, 2263 
\bibitem[Lauer et al.(2012)]{lau12} Lauer, T.~R., Bender, R., 
Kormendy, J., Rosenfield, P., \& Green, R.~F.\ 2012, \apj, 745, 121
\bibitem[Lewis et al.(2015)]{lew15} Lewis, A.~R., Dolphin, 
A.~E., Dalcanton, J.~J., et al.\ 2015, arXiv:1504.03338  
\bibitem[Li, Wang \& Wakker(2009)]{li09} Li, Z., Wang, Q.~D., 
\& Wakker, B.~P.\ 2009, \mnras, 397, 148 
\bibitem[Li et al.(2011)]{li11} Li, Z., Garcia, M.~R., 
Forman, W.~R., et al.\ 2011, \apjl, 728, L10  
\bibitem[MacArthur et al.(2009)]{mac09} MacArthur, L.~A., 
Gonz{\'a}lez, J.~J., \& Courteau, S.\ 2009, \mnras, 395, 28 
\bibitem[Markwardt(2009)]{mar09} Markwardt, C. B., 2009, ASPC, 411, 251M
\bibitem[Minchev et 
al.(2012)]{min12} Minchev, I., Famaey, B., Quillen, A.~C., et al.\ 2012, \aap, 548, AA126 
\bibitem[McConnachie et al.(2005)]{mcc05} McConnachie, A.~W., 
Irwin, M.~J., Ferguson, A.~M.~N., et al.\ 2005, \mnras, 356, 979 
\bibitem[Morris 
\& Serabyn(1996)]{mor96} Morris, M., \& Serabyn, E.\ 1996, \araa, 34, 645 
\bibitem[Melchior 
\& Combes(2013)]{mel13} Melchior, A.-L., \& Combes, F.\ 2013, \aap, 549, A27 
\bibitem[O'Connell(1999)]{oco99} O'Connell, R.~W.\ 1999, \araa, 37, 603 
\bibitem[Olsen et al.(2006)]{ols06} Olsen, K.~A.~G., Blum, 
R.~D., Stephens, A.~W., et al.\ 2006, \aj, 132, 271 
\bibitem[Rosenfield et al.(2012)]{ros12} Rosenfield, P., 
Johnson, L.~C., Girardi, L., et al.\ 2012, \apj, 755, 131
\bibitem[Saglia et 
al.(2010)]{sag10} Saglia, R.~P., Fabricius, M., Bender, R., et al.\ 2010, \aap, 509, A61 
\bibitem[Sarajedini 
\& Jablonka(2005)]{sar05} Sarajedini, A., \& Jablonka, P.\ 2005, \aj, 130, 1627 
\bibitem[Schlafly 
\& Finkbeiner(2011)]{sch11} Schlafly, E.~F., \& Finkbeiner, D.~P.\ 2011, \apj, 737, 103 
\bibitem[Skrutskie et al.(2006)]{skr06}Skrutskie, M. F., Cutri, R. M.,
  Stiening, R., Weinberg, M. D., Schneider, S., Carpenter, J. M.,
  Beichman, C., Capps, R., Chester, T., Elias, J. et al., 2006, AJ,
  131, 1163S
\bibitem[Stephens et al.(2003)]{ste03} Stephens, A.~W., 
Frogel, J.~A., DePoy, D.~L., et al.\ 2003, \aj, 125, 2473 
\bibitem[Thilker et al.(2005)]{thi05} Thilker, D.~A., Hoopes, 
C.~G., Bianchi, L., et al.\ 2005, \apjl, 619, L67 
\bibitem[V{\'a}zquez 
\& Leitherer(2005)]{vaz05} V{\'a}zquez, G.~A., \& Leitherer, C.\ 2005, \apj, 621, 695
\bibitem[Williams et al.(2014)]{wil14} Williams, B.~F., Lang, 
D., Dalcanton, J.~J., et al.\ 2014, \apjs, 215, 9 
\bibitem[Williams et al.(2005)]{wil05} Williams, B.~F., 
Garcia, M.~R., Kong, A.~K.~H., Primini, F.~A., 
\& Murray, S.~S.\ 2005, \apj, 620, 723 
\end{thebibliography}
\end{document}